\documentclass[useAMS,usenatbib]{mnras}

\usepackage{aas_macros}
\usepackage{graphicx}
\usepackage{hyperref}
\usepackage{pdflscape}
\usepackage{ wasysym }
\usepackage{txfonts}
\usepackage{natbib}
\usepackage{hyperref}
\usepackage{xspace}
\usepackage{lineno}
\usepackage{rotating}
\usepackage{pdflscape}
\newcommand{\uvot}{\textit{Swift}/UVOT}
\newcommand{\xrt}{\textit{Swift}/XRT}

\newcommand{\nus}{NuSTAR}

\newcommand{\po}{power-law}

\newcommand{\lp}{log-parabola}

\newcommand{\one}{1ES\,0229+200}
\newcommand{\itu}{infrared-to-ultraviolet}
\newcommand{\nh}{N$_H$}
\newcommand{\ch}{$\chi_{red}^2$}

\title[Constraining X-ray emission in  HBL blazars using multiwavelength observations]{Constraining X-ray emission in  HBL blazars using multiwavelength observations }

\author[A. Wierzcholska, S.Wagner]{
    Alicja Wierzcholska$^{1,2}$\thanks{E-mail: alicja.wierzcholska@ifj.edu.pl}, Stefan J. Wagner$^{1}$\\
    $^{1}$Landessternwarte, Universit\"at Heidelberg, K\"onigstuhl 12, D 69117 Heidelberg, Germany \\
    $^{2}$Insitute of Nuclear Physics, Polish Academy of Sciences, ul. Radzikowskiego 152, 31-342 Krak\'{o}w, Poland\\
   }

\begin{document}

\pagerange{\pageref{firstpage}--\pageref{lastpage}} \pubyear{2020}
\maketitle
\label{firstpage}

\begin{abstract}
The X-ray spectrum of extreme HBL type blazars is located in the synchrotron branch of the broadband spectral energy distribution (SED), at energies below the peak. 
A joint fit of the extrapolated X-ray spectra together with a host galaxy template allows characterizing the synchrotron branch in the SED. 
 The X-ray spectrum is usually characterized either with a  pure or a curved power-law model. 
In the latter case, however,  it is hard to distinguish an intrinsic curvature from excess absorption. 
In this paper, we focus on five well-observed  blazars: 1ES\,0229+200, PKS\,0548-322, RX\,J1136+6737, 1ES\,1741+196, 1ES\,2344+514. 
We constrain the infrared-to-X-ray emission of these five blazars using a model that is characterized by the host galaxy, spectral curvature, absorption, and ultraviolet excess to separate these spectral features. 
 In the case of four sources: 1ES\,0229+200, PKS\,0548-322,  1ES\,1741+196, 1ES\,2344+514 the spectral fit with the atomic neutral hydrogen from the Leiden Argentina Bonn Survey result in a significant UV excess present in the broadband spectral energy distribution. 
Such excess can be interpreted as an additional component, for example, a blue bump.
However, in order to describe spectra of these blazars without such excess, additional absorption to the atomic neutral hydrogen from the Leiden Argentina Bonn Survey is needed.

\end{abstract}


\section{Introduction}\label{intro}
The class of blazars is characterized by a polarized and highly variable non-thermal continuum emission and composed of BL Lacertae (BL Lac) type sources and flat spectrum radio quasars (FSRQs).
The distinction has  historically been  defined based on the equivalent width of the optical emission lines.
According to the unified model \citep[e.g.][]{Urry95}, blazars are  observed at small angles of the observer's line of sight to the relativistic jet axis \citep[e.g.][]{begelman84}.
The electromagnetic radiation that is emitted by blazars is observed in the full energy range, starting from radio frequencies up to very high energy $\gamma$ rays \citep[e.g.][]{Wagner2009, Abramowski2014}.
Temporal variability, observed in all energy regimes on different time scales from minutes up to years is a characteristic feature of these sources \citep[e.g.][]{2155flare, Wierzcholska_0048, Liao15, Wierzcholskas5}.
Furthermore, temporal flux changes in blazars are often associated with spectral variability \citep[e.g.][]{Xue06, Bottcher10, Wierzcholska2016, Siejkowski_2017}.

A typical spectral energy distribution (SED) of blazars  is described by two broad emission components. 
The  low-energy peak in the SED is usually explained by synchrotron radiation of relativistic electrons from the jet, 
while the  high-energy component can be explained in leptonic or hadronic scenarios \citep[see, e.g.][]{Dermer92, Sikora94, Mucke13, Bottcher13}.
In the case of leptonic models, the high-energy peak can be either caused by inverse Compton scattering of relativistic electrons from the jet (synchrotron-self-Compton models, SSC) or by photon fields external to the jets (external Compton models, EC). 
The external field of photons can be caused by emission observed from  a dusty torus or broad line region. 

The position of two peaks allows us to distinguish three subgroups of BL Lac type blazars, namely: high-, intermediate- and low-energy  peaked BL Lac objects: HBL, IBL, LBL, respectively \citep[see, e.g.][]{padovani95, fossati98, Abdo2010}.
In the case of HBL type blazars the synchrotron peak is located in the X-ray domain ($\nu_{s}>10^{15}$\,Hz) \citep[][]{Abdo2010}.

Two functions are commonly used to describe the synchrotron spectrum of blazars. 
These are a single power-law model and a curved power-law (also known as logparabola). 
The first one is characterized by a photon index $\gamma$, which is related to the index of input particles as $q = 2\gamma +1$ \citep{Rybicki}.
Similarly, a curved spectrum can be produced by logparabolic particle distribution \citep{Paggi09} and is an indication for statistical acceleration process \citep[e.g.][]{Massaro2004, Massaro2006, Massaro2008, Tramacere2007}.

The emission observed in the soft X-ray range is affected by absorption of the interstellar medium,  along line of sight, in our Galaxy. 
The effect of the absorption  extends up to about 10\,keV with the most substantial influence up to 2\,keV
and it is calculated using a neutral hydrogen column density.
The hydrogen column density observations have been reported in different surveys. 
The  Leiden Argentine Bonn Survey \citep[LAB, ][]{Kalberla2005}  includes only N$_{HI}$ measure. 


\cite{Willingale13} have proposed a  measure of N$_{H,tot}$, which includes both the atomic gas column density N$_{HI}$ and molecular hydrogen column density N$_{H_2}$.
The N$_{HI}$ value is taken from the LAB survey, while  N$_{H_2}$ is estimated based on maps of infrared dust emission and dust-gas ratio as provided by \cite{Schlegel98} and \cite{Dame01}, respectively.
As an alternative, spectra of blazars can be fitted with a free value of column density, which allows  the derivation of the total X-ray absorption \citep[e.g.][]{Furniss2013, Wierzcholska2016, Gaur18}.

The X-ray spectra alone cover too small range in energy to decompile the effect of absorption from the range of different effects that can cause spectral curvature. 
Thus a more in-depth look into multifrequency observations of a source is needed.

In this paper, we investigate X-ray observations of HBL type blazars together with multiwavelength data  of synchrotron part of the spectrum and a host galaxy template to study properties of the X-ray spectra and host galaxy emission observed in blazars selected. 
The paper is organized as follows: Sect.~\ref{sample} introduces the sample of targets studied, Sect.~\ref{data} presents detail on data presented in the paper and data analysis detail, Sect.~\ref{results} shows the results.
The work is summarized in Sect.~\ref{summary}. 

\noindent

\section{The sample} \label{sample}
52 HBL type blazars have been detected in the TeV energy range (TeVCat\footnote{\url{http://tevcat.uchicago.edu/}}).
Most of them were observed with \xrt\ with PC or WT mode of the instrument \citep[see for detail][]{Wierzcholska2016}.
In this work, we aim to characterize emission observed from infrared to X-ray range for extreme HBL type blazars. 
We note here that extreme blazars are described by extremely energetic synchrotron emission, and the inverse Compton bump extends to the very high energy $\gamma$-ray regime. 
For such sources, the X-ray spectrum is located at energies below the peak of the low energy bump in SED.
For further studies, we select sources, TeV $\gamma$-ray emitting blazars, which fulfil the following criteria:
\begin{itemize}
 \item the source belongs to HBL type blazars with the X-ray spectrum located at the growing part of the synchrotron bump in the SED,
  \item multiwavelength data covering low energy bump in the SED are available;
 \item archival observations of the global SED (e.g., taken from ASDC\footnote{\url{http://www.asdc.asi.it/}})  revealed prominent emission visible in the \itu\ range indicating that host galaxy emission is significantly stronger than emission originating from all other components.
\end{itemize}
As a result the selected sample consists of 5 sources: 1ES\,0229+200, PKS\,0548-322, RX\,J1136+6737, 1ES\,1741+196, 1ES\,2344+514 as targets for further studies.\footnote{We note here also that during the review process of the paper, the MAGIC Collaboration reported the discovery of a new sample of hard-TeV extreme blazars \citep{MAGIC_extreme}. These AGNs are also possible targets: TXS\,0210+515, RBS\,0723, 1ES\,1426+428, 1ES\,2037+521, and RGB\,J2042+244, for similar studies as presented in the paper.}

Two potential sources: Mrk\,421, and Mrk\,501 are not included in these studies due to significant X-ray variability seen in the X-ray observations and the fact that most \xrt\ data taken for these blazars are in WT mode. 
We limit our analysis to PC mode data only since data  taken in the WT mode are affected by charge redistribution problems inherently related to how the CCD of the instrument is read \citep[also noted by, e.g.][]{Massaro2008, Wierzcholska2016}. This causes that in the case of the longterm integrated X-ray spectra, wavy features can be present.

\section{Data and  analysis methods} \label{data}

In order to ensure good data coverage of the low energy bump in the SEDs of  sources studied: 1ES\,0229+200, PKS\,0548-322, RX\,J1136+6737, 1ES\,1741+196, 1ES\,2344+514
multiwavelength data are used. In the case of all blazars, WISE, 2MASS, ATOM, \uvot, \xrt\ data are used. In the case of 1ES\,0229+200 also \nus\ observations are studied.

\subsection{Data analysis}

\subsection*{NuSTAR data}

The Nuclear Spectroscopic Telescope Array (NuSTAR) is a satellite instrument dedicated to observations in the hard X-ray regime of 3-79 keV \citep{Harrison2013}.
Out of the five targets selected for these studies, only \one\ was observed with \nus. Details on three pointings performed are summarized in Table\,\ref{table:nustar}.
All observations were performed in the \verb|SCIENCE| mode.

The raw data were processed with \nus\ Data Analysis Software package
(\verb|NuSTARDAS|, 
released as part of
\verb|HEASOFT|~6.25)
using standard 
\verb|nupipeline|
task. 
Instrumental response matrices and effective area files were produced with
\verb|nuproducts|
procedure. 
The spectral analysis was performed for the channels corresponding to the energy band of 3-79\,keV. 
We mention here that \nus\ observations of \one\ have been previously reported by \cite{Bhatta_0229} and \cite{Pandey_0229}.
However, \cite{Bhatta_0229} have discussed only one \nus\ observation of \one, while \cite{Pandey_0229} focused on temporal variability.

\subsection*{\textit{Swift}-XRT data}
The X-ray data used in this work are the ones collected with Swift \citep{Gehrels}, which are exactly the same sets as studied by \cite{Wierzcholska2016}, reanalyzed with a more recent version of the HEASoft software (6.25) with the recent CALDB. 
All events were cleaned and calibrated using \verb|xrtpipeline| task, and the PC mode data in the energy range of 0.3-10\,keV with grades 0-12 were analyzed.
The reanalyzed spectra are entirely consistent with the previous ones.

 \subsection*{\textit{Swift}-UVOT data}
The UVOT instrument onboard \textit{Swift}  measures the ultraviolet and optical emission simultaneously with the X-ray telescope.
The observations are taken in the UV and optical  bands with 
the central wavelengths of UVW2 (188 nm), UVM2 (217 nm), UVW1 (251 nm), U (345 nm), B
(439 nm), and V (544 nm). 
The instrumental magnitudes were calculated using \verb|uvotsource| including all photons from a circular region with radius 5''.
The background was determined from a circular region with a radius of 10'' near the source region, not contaminated with the signal from nearby sources. 
The flux conversion factors as provided by \cite{Poole08} were used. 
All data were corrected for the dust absorption using the reddening $E(B-V)$   as provided by \cite{Schlafly} and summarized in Table\,\ref{table:datasummary}. 
The ratios of the extinction to reddening, $A_{\lambda} / E(B-V)$, for each filter were provided by \cite{Giommi06}.
The reddening $E(B-V$) and atomic hydrogen column density (\nh) are related with the following formula \nh = 8.3 $\cdot 10^{21}E(B-V)$ \citep{Liszt2014}.

\noindent

\subsection*{ATOM data}
ATOM (Automatic Telescope for Optical Monitoring) is a 75\,cm optical telescope located in Namibia near the H.E.S.S. site \citep{Hauser2004}.
The instrument measures optical magnitudes in four optical filters: B (440 nm), V (550 nm), R (640 nm), and I (790 nm).

For all sources, the data collected in the period starting from 2007 to 2015 have been analyzed using an aperture of 4'' radius and differential photometry. The observations have been corrected for dust absorption.

ATOM observations are available for the following sources from the sample, namely: 1ES\,0229+200,  PKS\,0548-322, and 1ES\,1741+196. In the case of each of the sources, there is no significant variability detected in the ATOM bands. 

\subsection*{2MASS data}
The 2MASS data are taken from the 2MASS All-Sky Point Source Catalog (PSC) \citep{TwoMassPSC}. 
The data were corrected for the dust absorption using the same $E(B-V)$ factors as for the optical data mentioned previously.

\subsection*{WISE data}
Wide-field Infrared Survey Explorer (WISE) is a space telescope which performs observations in the infrared energy band at four wavelengths: 3.4\,$\mu$m (W1), 4.6\,$\mu$m (W2), 12\,$\mu$m (W3) and 22\,$\mu$m (W4). The spectral data were taken from the AllWISE Source Catalog and the light curve from the AllWISE Multiepoch Photometry Table\footnote{\url{http://wise2.ipac.caltech.edu/docs/release/allwise/}}. The magnitudes were converted to flux by applying the standard procedure \citep{Wright_2010}. For W1 and W2, we applied the colour correction of a power-law with index zero, as suggested for  Galactic emission. Since W3 and W4 are widely dominated by the non-thermal radiation and show similar flux densities, we used the colour correction for a power-law of index -2.
For none of the blazars studied, the infrared light curves does not show any  variability within uncertainties of the measurements.

\subsection{Simultaneity of the data}

Blazars are known for their variability observed in all energy regimes. 
Thus, simultaneous multiwavelength observations are essential to study broadband emission observed from blazars.
Unfortunately, single \xrt\ observations usually do not allow to constrain spectral parameters with small uncertainties.
In order to find a sustainable solution for this, here, we study  integrated X-ray spectra of five blazars.
However, only PC mode data is used, since high state observations are usually performed with the WT mode of \xrt.
In order to check if the effect of variability is significant, we quantified it by fitting a constant function to the light curve data points. 
For all cases, such a fit yields $\chi_{red}^2$ of 0.8-1.4,  showing no indication for variability in the X-ray regime for any of the targets.

Furthermore, thanks to simultaneous observations with \uvot, optical, and ultraviolet data collected with this instrument are averaged over the same period as X-ray data. 
Optical observations collected with ATOM are averaged over a similar period as X-ray monitoring data. 
We note here, that in the case of X-ray spectra obtained with \nus, exactly simultaneous \xrt\ and \uvot\ data are used.

\subsection{Spectral analysis and fitting procedure} \label{fitting}
In order to characterize X-ray spectra of five sources from the sample two different models, the absorbed power-law (\texttt{wabs\,$*$\,powerlaw} in \verb|XSPEC|) and absorbed logparabola (\texttt{wabs\,$*$\,logparabola} in \verb|XSPEC|) were tested.

A single power-law is  defined as:
\begin{equation}
 \frac{dN}{dE}=N_p  \left( \frac{E}{E_0}\right)^{-{\gamma}},
\end{equation}
with the normalization $N_p$, and the photon index $\gamma$;
while a logparabola model is defined as:
\begin{equation}
 \frac{dN}{dE}=N_l  \left( \frac{E}{E_0}\right)^{-({\alpha+\beta \log (E/E_0)})},
\end{equation}
 with the normalization $N_l$, the spectral parameter $\alpha$ and the curvature parameter $\beta$. 
 Galactic absorption is a function of exponent:  $ e^{-(N_{H}^{S}+N_{H}^{A})\sigma (E)}$, where  N$_{H}^{S}$ is the hydrogen column density value taken for a given survey,  N$_{H}^{A}$ is the additional, intrinsic column density  and $\sigma$(E) is the non-Thomson  energy dependent photoelectric  cross section \citep{Morrison83}.
 For the case when there is an assumption of lack of an additional absorption, only N$_{H}^{S}$ is taken into account, and N$_{H}^{A}$ is fixed to zero. 

We note here that except for absorbing model described with \texttt{wabs}, we also tested \texttt{phabs} and \texttt{tbabs} \citep{wabsy}.
This resulted in similar spectral parameters, as presented in the paper and can be found in the online material to the paper.

\subsection{Fitting of different energy bands}
The fitting of the X-ray spectrum for \xrt\ observations is performed in two different energy bands: 0.3-10\,keV and 2.0-10\,keV. 
The energy band of 0.3-10\,keV includes complete spectral information from \xrt\ observations.
In this case, different corrections for the Galactic absorption are tested. 
The 2.0-10\,keV band has limited coverage and is slightly less affected by absorption. 
This is simply because the Galactic absorption plays a substantial role from low energies up to about 2\,keV \citep[e.g.][]{Campana2014} and thus by limiting the fitting range the influence of the Galactic absorption is marginal.

\subsection{Fitting of host galaxy}
The fitting procedure of a host galaxy profile to the observational data is the following:
\begin{enumerate}
\item The preferred X-ray spectral model, according to X-ray data analysis as presented by \cite{Wierzcholska2016}, is chosen for a  given object, as a starting point. 
\item The X-ray spectrum is then extrapolated to the lower energies, up to the IR regime, and added to the synthetic host galaxy profile.
\item The total profile is then fitted to the observational data points.
\end{enumerate}

The synthetic host galaxy profiles are generated with GRASIL \citep[for more details see][]{Silva98, Silvathesis, grasil}, a code to compute spectral evolution of stellar systems.
For this work,  the simulations of the standard model of an elliptical galaxy based on the model described in the manual pages\footnote{\url{http://adlibitum.oats.inaf.it/silva/grasil/modelling/modlib.html}} is used. 
We generated $11 \times 26$ synthetic spectra for a set of 11 different infall masses, $M_\textrm{inf}$: $10^7$, $10^{7.5}$, ... , $10^{12}$ $M_\odot$, and 26 different galaxy ages, $t_\textrm{gal}$:  $0.1$, $0.6$, ..., $13.1$ Gyr.

The fitting method is as follows:
\begin{enumerate}
 \item The age of the galaxy is set to a given value.
 \item The $\texttt{scipy.interpolate.interp2d}$ function \citep{SciPy} is used to interpolate generated spectra (points) in two dimensions of infall mass and frequency in order to have a well-covered grid. 
 This returns a function $S_\textrm{syn}(\nu,M_\textrm{inf})$ which computes the spectral points for a given frequency and infall mass.
 \item the $S_\textrm{syn}$  (with added extrapolated X-ray spectrum) is fitted to the observational data using  $\texttt{scipy.optimize.curve\_fit}$ function.
 \item The procedure is repeated for each galaxy age.
 \item The best fit is selected according to the $\chi^2$ value.
\end{enumerate}

\noindent

\begin{table} 
\centering 
\begin{tabular}{c|c|c}
\hline
\hline
 Source &  Swift obsIDs  & $E(B-V)$   \\
(1) &  (2)& (3)     \\
\hline
1ES\,0229+200 & 00031249001-00031249050   &  0.1172  \\
PKS\,0548-322 & 00044002001-00044002065  &  0.0304   \\
RXJ\,1136+6737 & 00037135001-00040562010  &  0.0074   \\
1ES\,1741+196  & 00030950001-00040639018   &  0.0759   \\
1ES\,2344+514 & 00035031001-00035031121  &  0.1819   \\
\hline
\end{tabular}
\caption[]{Details on data studied. The following columns present: (1) name of the source; (2) Swift (XRT and UVOT) observation IDs used in the paper; (3) $E(B-V)$ reddening coefficient.}
  \label{table:datasummary}
\end{table}

\section{Results} \label{results}

\subsubsection*{1ES\,0229+200}
The studies of \one\ performed by \cite{Kaufmann2011} revealed unusual spectral characteristics of the X-ray emission detected up to about 100\,keV without any significant cut-off in the energy range studied. 
The model preferred by the authors has been a single \po, however, with an indication of excess absorption above the Galactic value. 
\cite{Kaufmann2011} has also shown that a cut-off characterizes the SED of \one\ in the low energy part of the synchrotron emission, located in the UV regime. 
In more recent studies, \cite{Wierzcholska2016} have demonstrated that the longterm integrated X-ray spectrum in the energy range of 0.3-10\,keV of \one\ could be described both with a single \po\ or a \lp\ model.
The authors have also shown that according to \ch\ values of the fits, the preferred \nh\ amount is the one from the LAB survey.
Within the uncertainties, this result is consistent with the work by \cite{Kaufmann2011}.

\begin{figure*}
\centering{\includegraphics[width=0.495\textwidth]{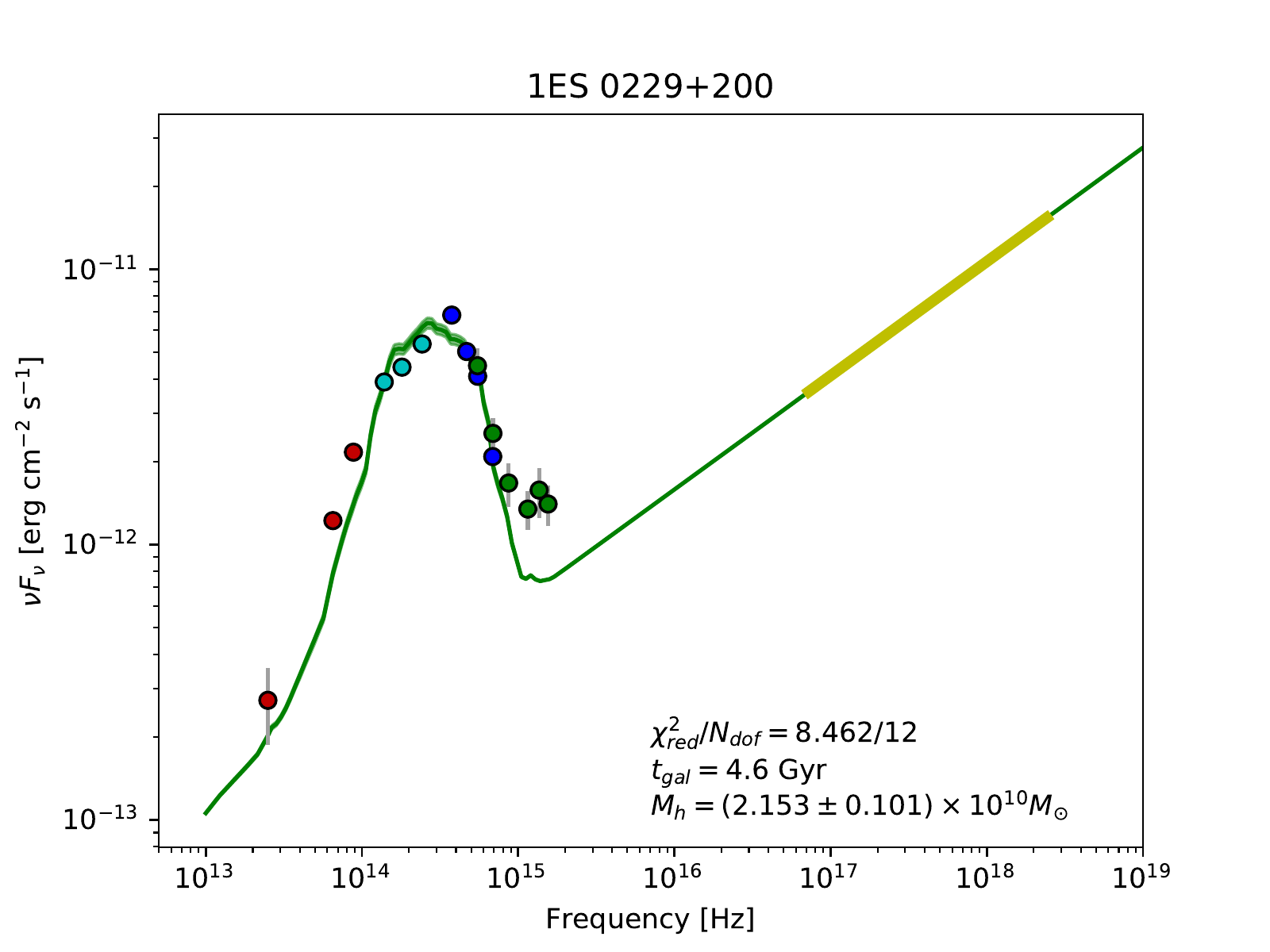}}
\centering{\includegraphics[width=0.495\textwidth]{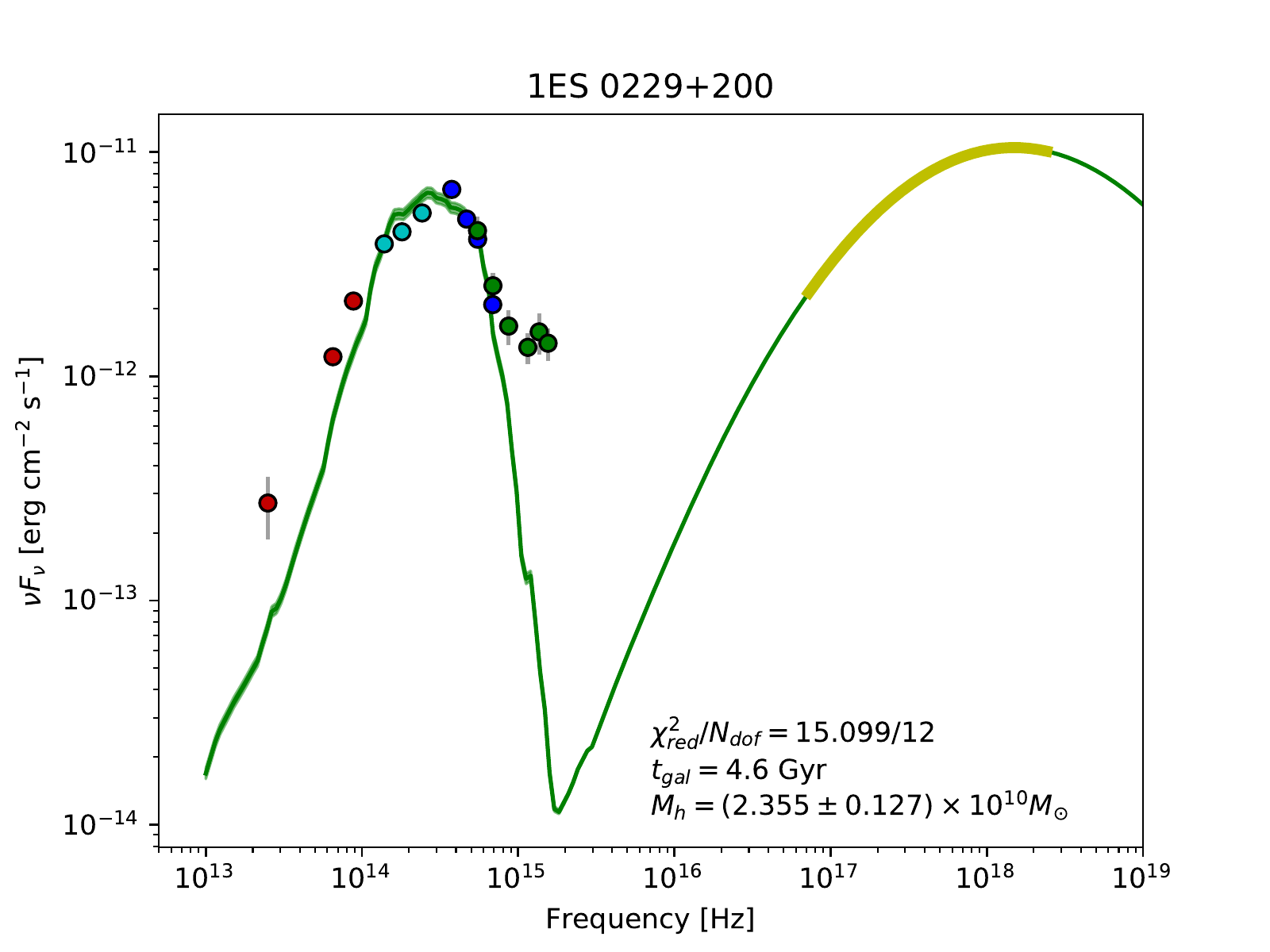}}
\caption[]{Broadband SED of 1ES\,0229+200. Left: modelling with X-ray spectrum fitted with the power-law model in the energy range of 0.3-10\,keV with \nh\ value taken from \cite{Kalberla2005}; right: same as left but X-ray spectrum fitted with the logparabola model. 
Red points present WISE data, light blue 2MASS data, dark blue ATOM data, and green Swift-UVOT.}
\label{figure:spectra_global}
\end{figure*}

In the first step, all X-ray observations (as analysed by \cite{Wierzcholska2016}) collected with \xrt\  during the period of monitoring in the energy range of 0.3-10\.keV were fitted with a single \po\ and a \lp\ model, in both cases with the Galactic absorption (\nh$=8.06$\,$\cdot$10$^{20}$cm$^{-2}$, \citealt{Kalberla2005}).
Fits parameters of both models are given in Table~\ref{table:results}. 
The figures illustrating fits with the data points are included in the online material. 
Both spectra are then used together with the host galaxy templates, as described in Sect.~\ref{fitting},  to fit infrared-to-ultraviolet observations of \one. 
The results of the fits are presented in Fig.~\ref{figure:spectra_global}. 
In the case of the \po\ spectral model used, a  fit to \itu\ data is characterized  with a significant excess  observed in the UV regime. 
 We note here, that, the UV excess is defined as a difference between model and fluxes observed in the UV range.

\begin{figure*}
\centering{\includegraphics[width=0.495\textwidth]{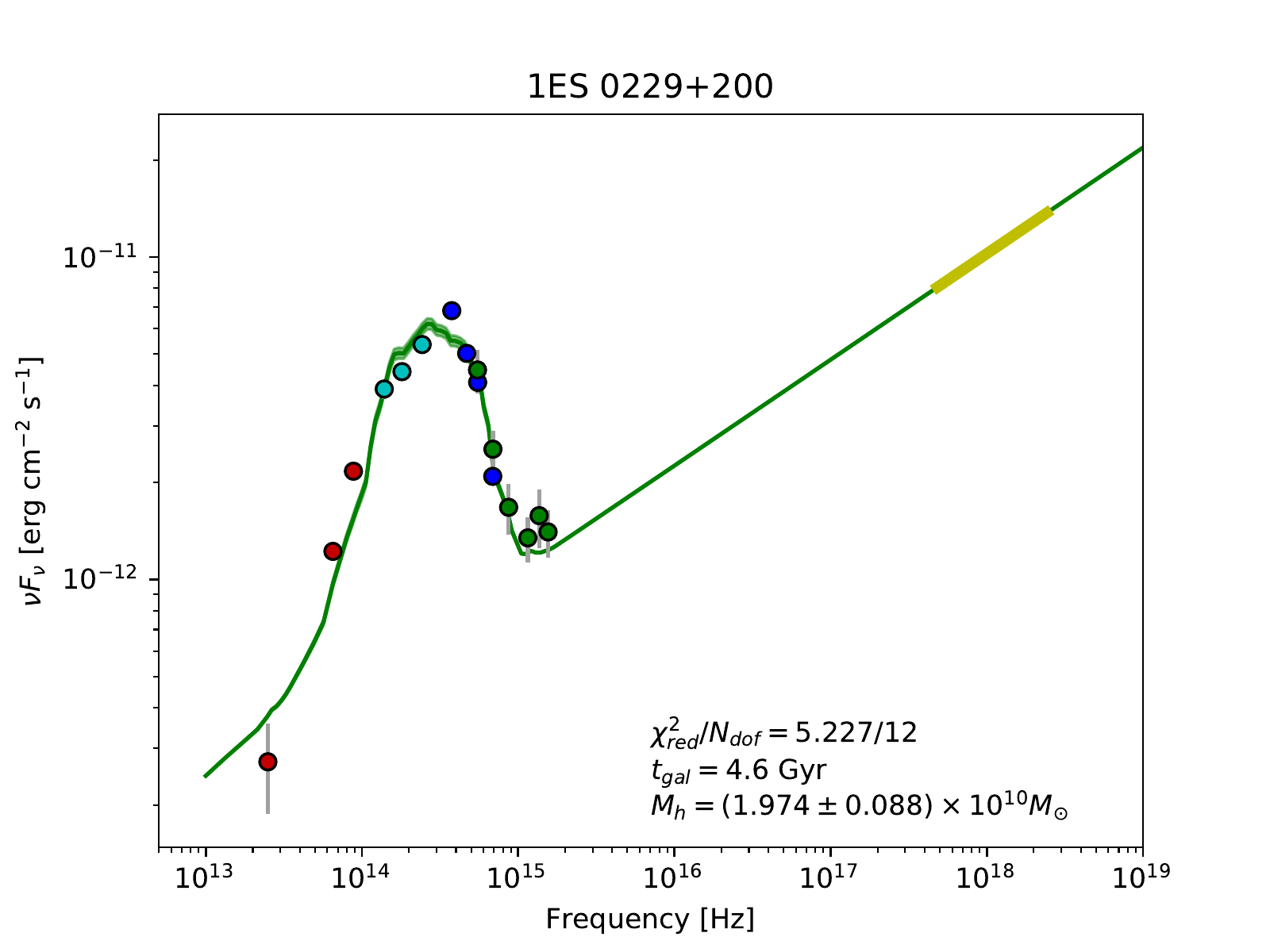}}
\centering{\includegraphics[width=0.495\textwidth]{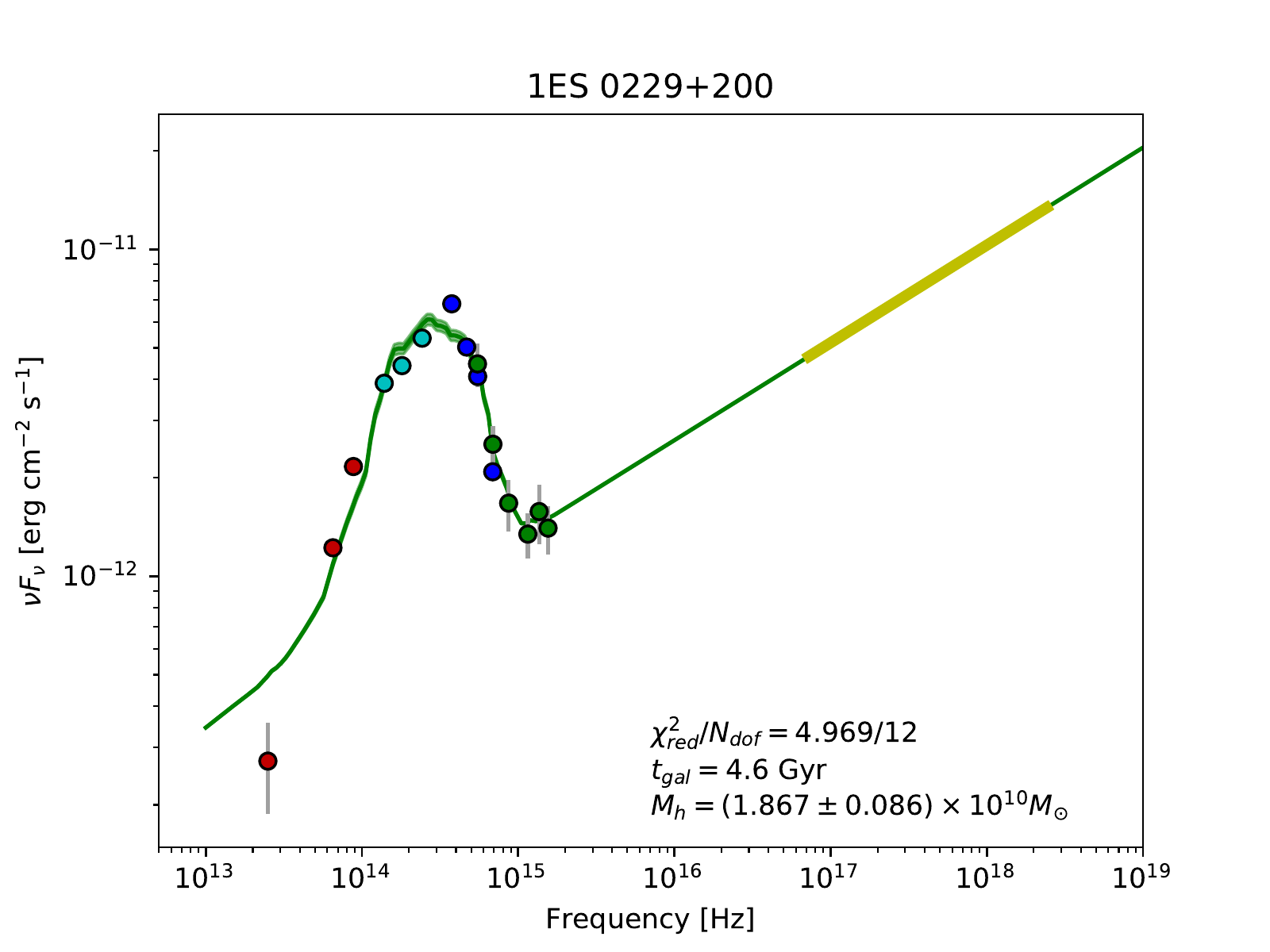}}

\caption[]{Broadband SED of 1ES\,0229+200. Left: modelling with X-ray spectrum fitted with the power-law model in the energy range of 2.0-10\,keV with \nh\ value taken from \cite{Kalberla2005}; right: modelling with X-ray spectrum fitted with the power-law model in the energy range of 0.3-10\,keV with \nh\ value taken from \cite{Willingale13}.
Points colour coding same as in Fig.\ref{figure:spectra_global}.}
\label{figure:spectra_limited}
\end{figure*}

In order to check how the excess observed depends on the Galactic absorption, the longterm integrated X-ray spectrum has been fitted again with the \po\ model in two different ways. 
\begin{itemize}
 \item In the first approach, the \nh\ value by \cite{Kalberla2005} is used, but the energy range is limited to 2.0-10\,keV. 
The fitting procedure used in this case is the same as described before, and the same multiwavelength data set is used. 
The resulting fit parameters are collected in the Table~\ref{table:results}, and the plot is presented in Fig.~\ref{figure:spectra_limited}.

\item 
In the second approach, the X-ray energy range used for the fitting is the entire \xrt\ energy range, but the fixed \nh\ value is higher than the one mentioned above.
 \cite{Willingale13} have suggested that in order to determine appropriately the influence of the Galactic absorption N$_{H,tot}$ value, which is a sum of the atomic gas column density N$_{HI}$ and the molecular hydrogen column density N$_{H_2}$ should be used.
Here, N$_{HI}$ is taken form the LAB survey \citep{Kalberla2005}, while N$_{H_2}$ is estimated using maps of dust infrared emission by \cite{Schlegel98} and the dust-gas ratio by \cite{Dame01}.
For the case of \one, N$_{H,tot}$  is equal to 11.80 $\cdot$ 10$^{20}$\,cm$^{-2}$. 
The fit to the data with the \po\ model plus the Galactic hydrogen absorption as provided by \cite{Willingale13} is presented in Fig.~\ref{figure:spectra_limited}
and fit parameters are collected in Table~\ref{table:results}.
\end{itemize}

In the case of the spectrum fitted to the limited energy range, the host galaxy is well described, and a UV excess is not required. 
The fit with the Galactic column density  value as provided by \cite{Willingale13} also does not require a UV excess and more significant discrepancy in the observed and modelled infrared emission in the low energy part of the WISE range. 

Previous works dedicated to X-ray studies of \one\ also revealed additional absorption observed in the case of \one. 
\cite{Kaufmann2011} have found that X-ray spectra of \one\ are well described with a single \po\ model with the additional absorption that can be either intrinsic to the blazar or in the line of sight to the observer or in the Milky Way. 
 
We note here also, that the evidence for UV excess in the X-ray spectrum of 1ES\,0229+200 has been reported by \cite{Costamante2018}.
The authors suggested that it can be either due to an additional emitting component or could be explained by thermal emission from the AGN. 
While no explanation could be identified, the shape of the X-ray spectrum required an excess emission in the UV range or excess absorption.

\subsubsection*{NuSTAR observations of \one}
In the case of one target from the sample, \one, hard X-ray data collected with \nus\ are available. 
This allows testing stability of the results obtained with \xrt, and check if additional X-ray observations ranging to 79\,keV improve the fit. 

In total, three \nus\ observations of \one\ were performed, all in October 2013.
During the \nus\ campaign, the source was also observed with \xrt\ and \uvot, providing almost a strictly simultaneous overlap of the broadband spectral energy distribution. 
Table~\ref{table:nustar} presents details on data that have been gathered with \nus\ and \xrt\ during \nus\ observation period. 

\begin{figure}
 \centering{\includegraphics[width=0.5\textwidth]{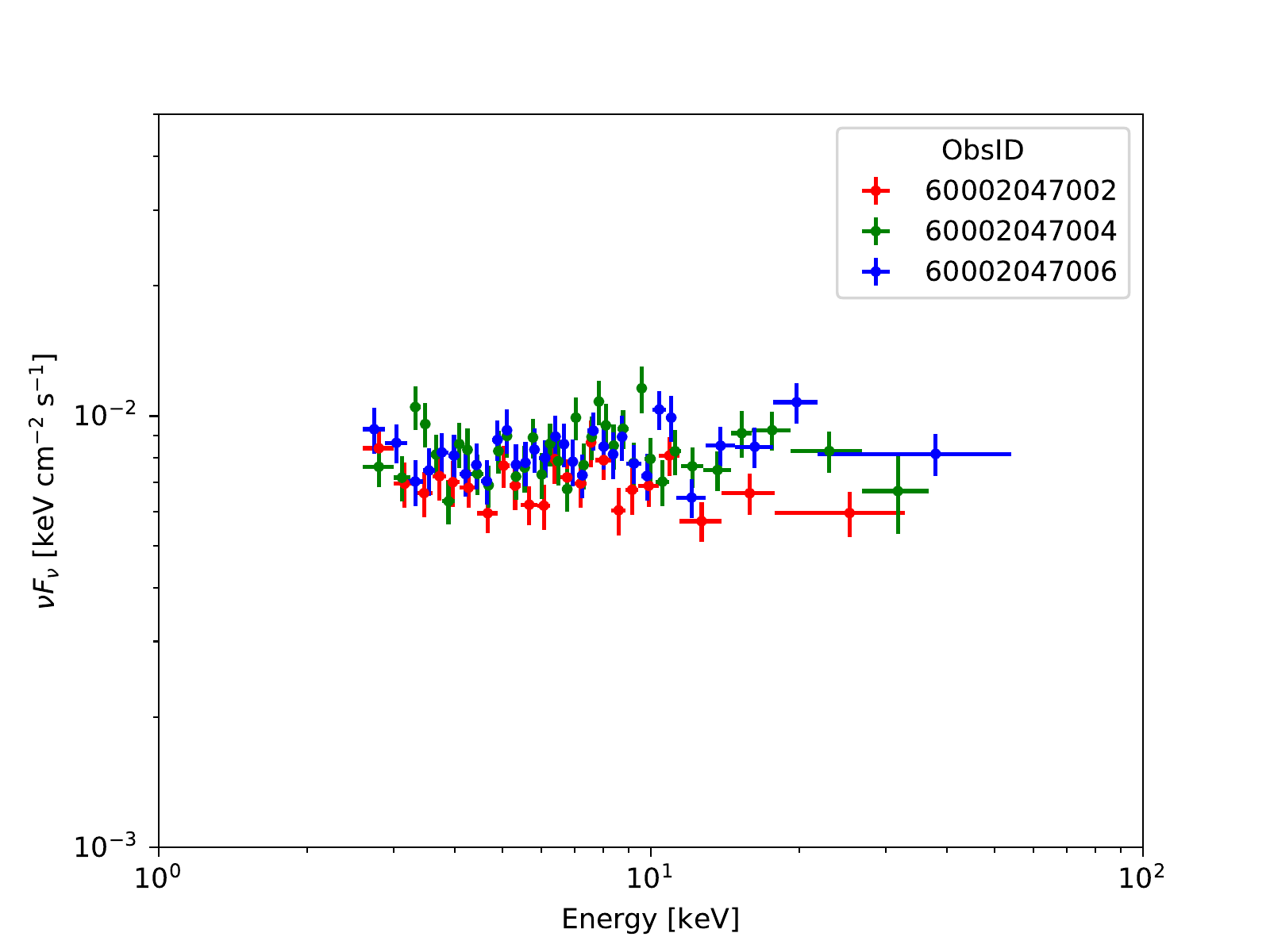}}
 \caption[]{The spectral energy distribution of \one\ as seen with \nus. Three different colors correspond to different ObsIds: nu60002047002, nu600020470042, and nu60002047006. For each ObsId, the data are fitted with a single \po\ model.  }
 \label{figure:nus_spec}
\end{figure}

\begin{figure*}
\centering{\includegraphics[width=0.3\textwidth]{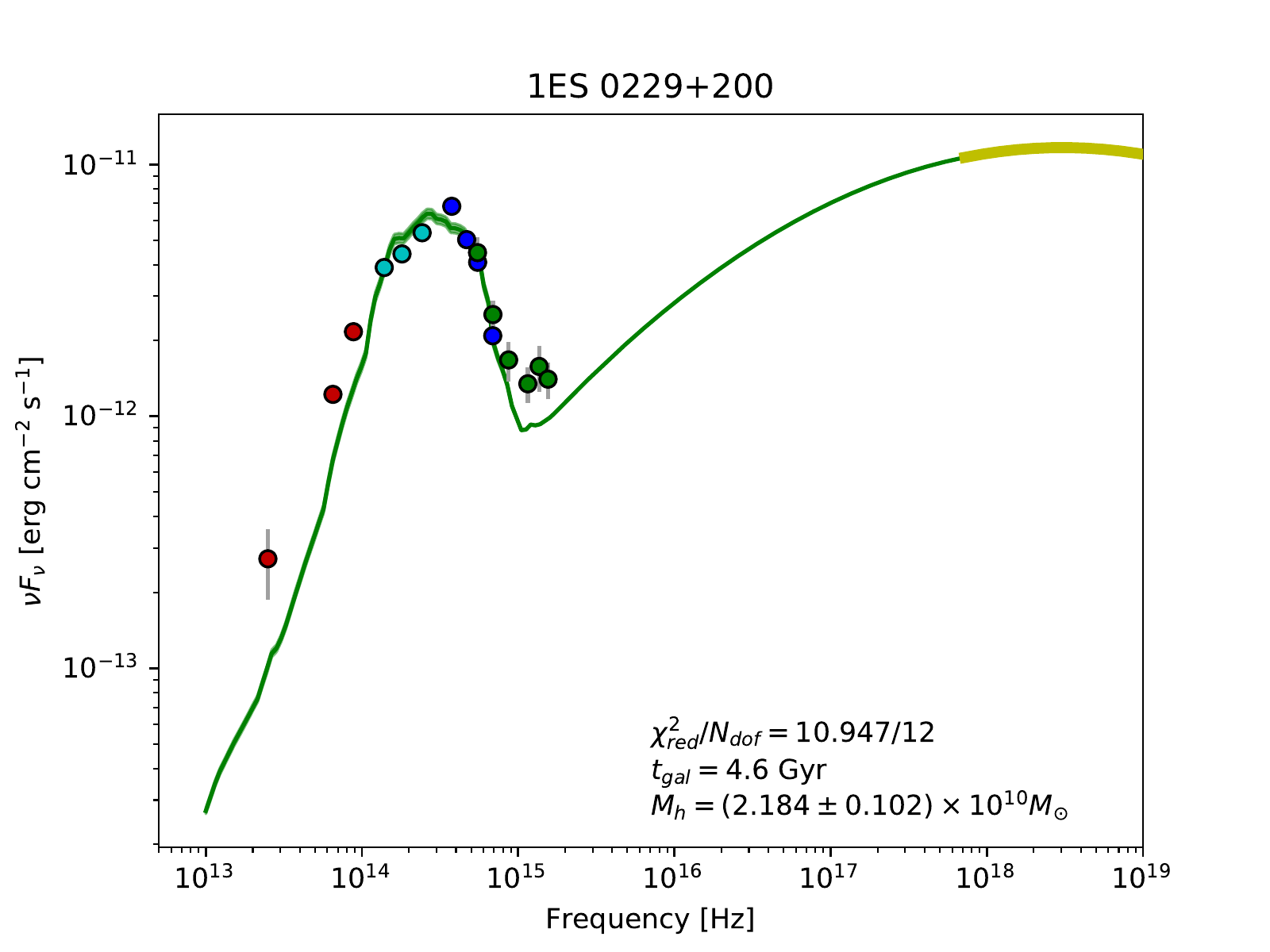}}
\centering{\includegraphics[width=0.3\textwidth]{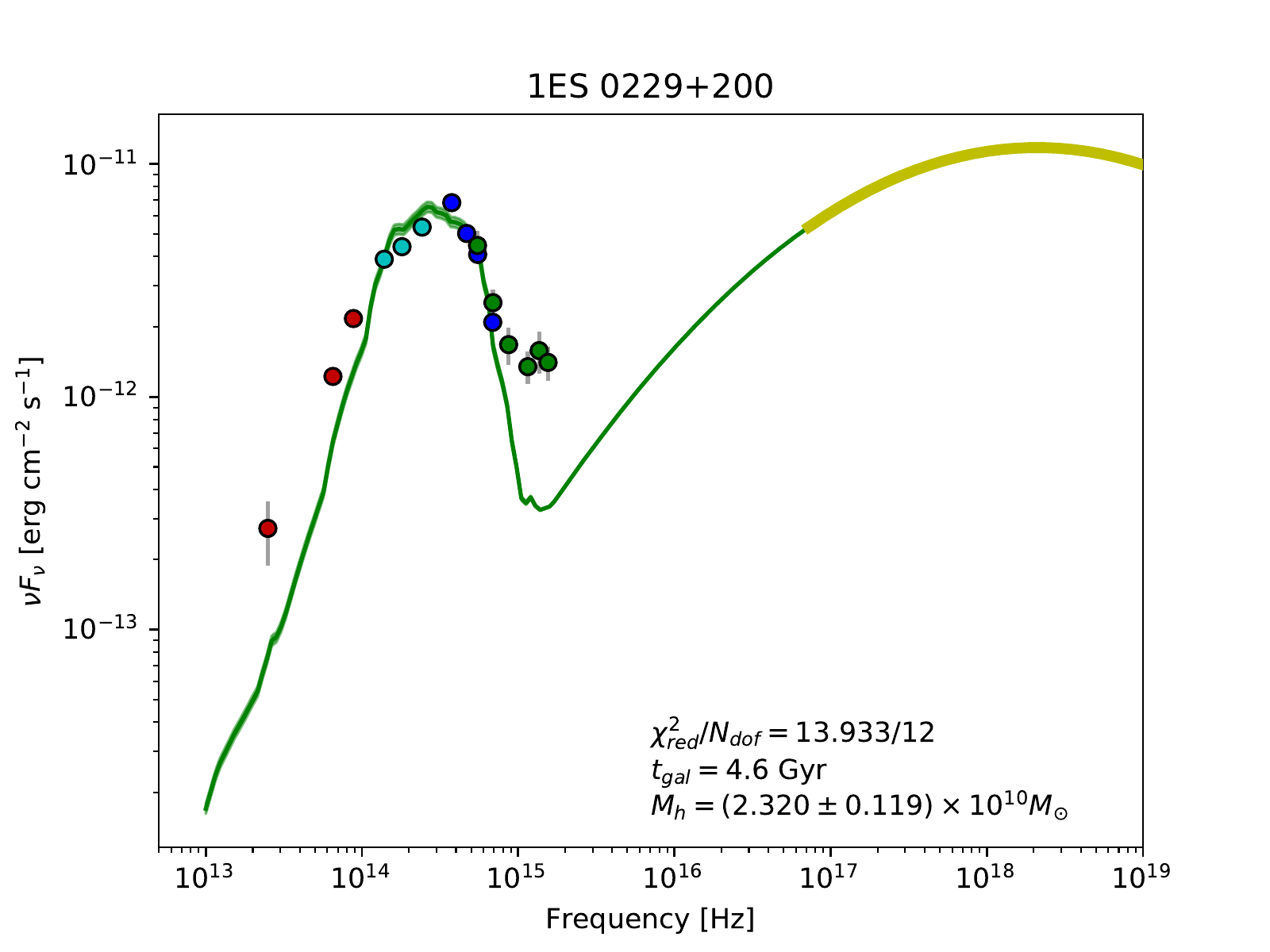}}
\centering{\includegraphics[width=0.3\textwidth]{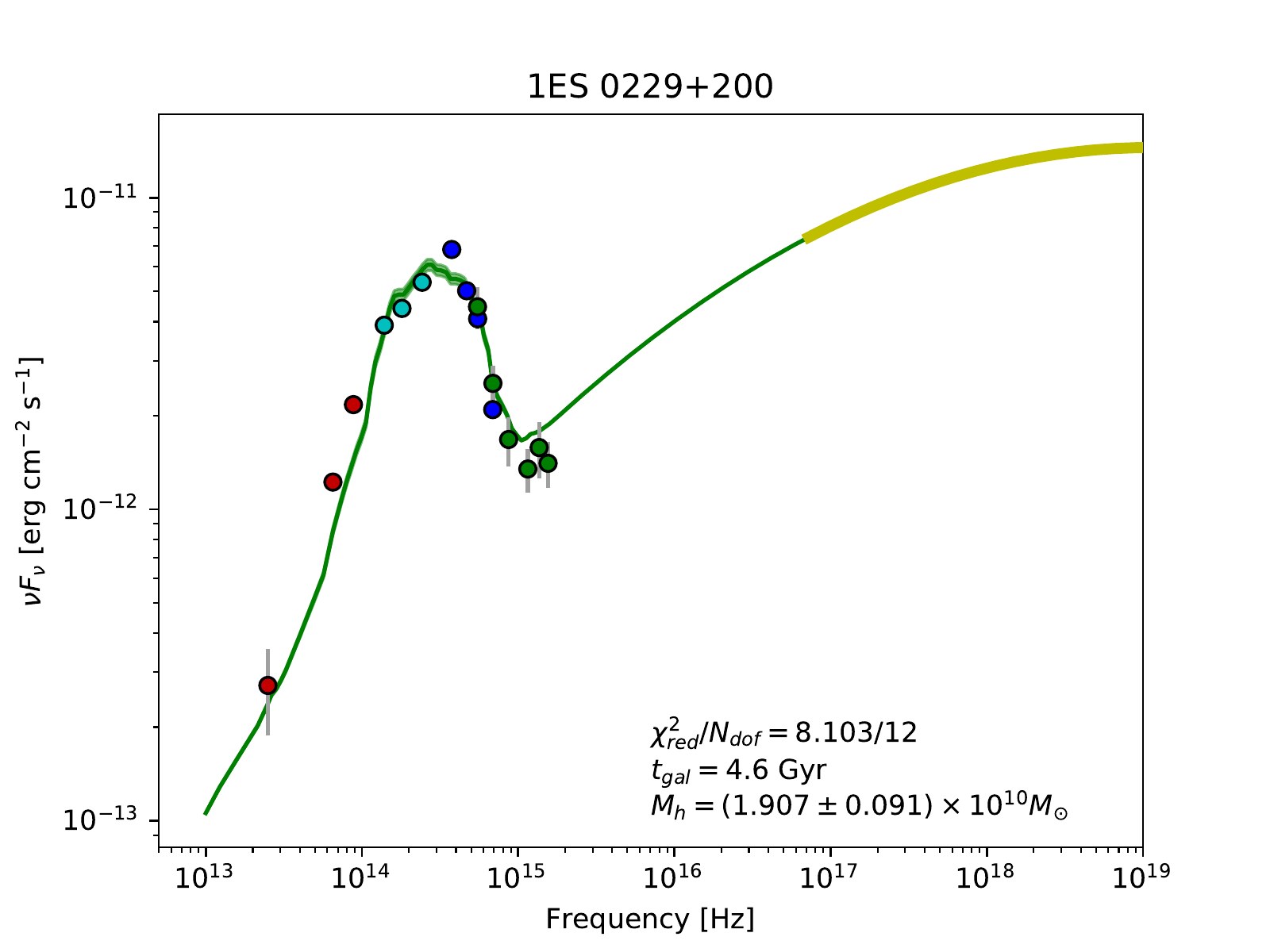}} \\
\centering{\includegraphics[width=0.3\textwidth]{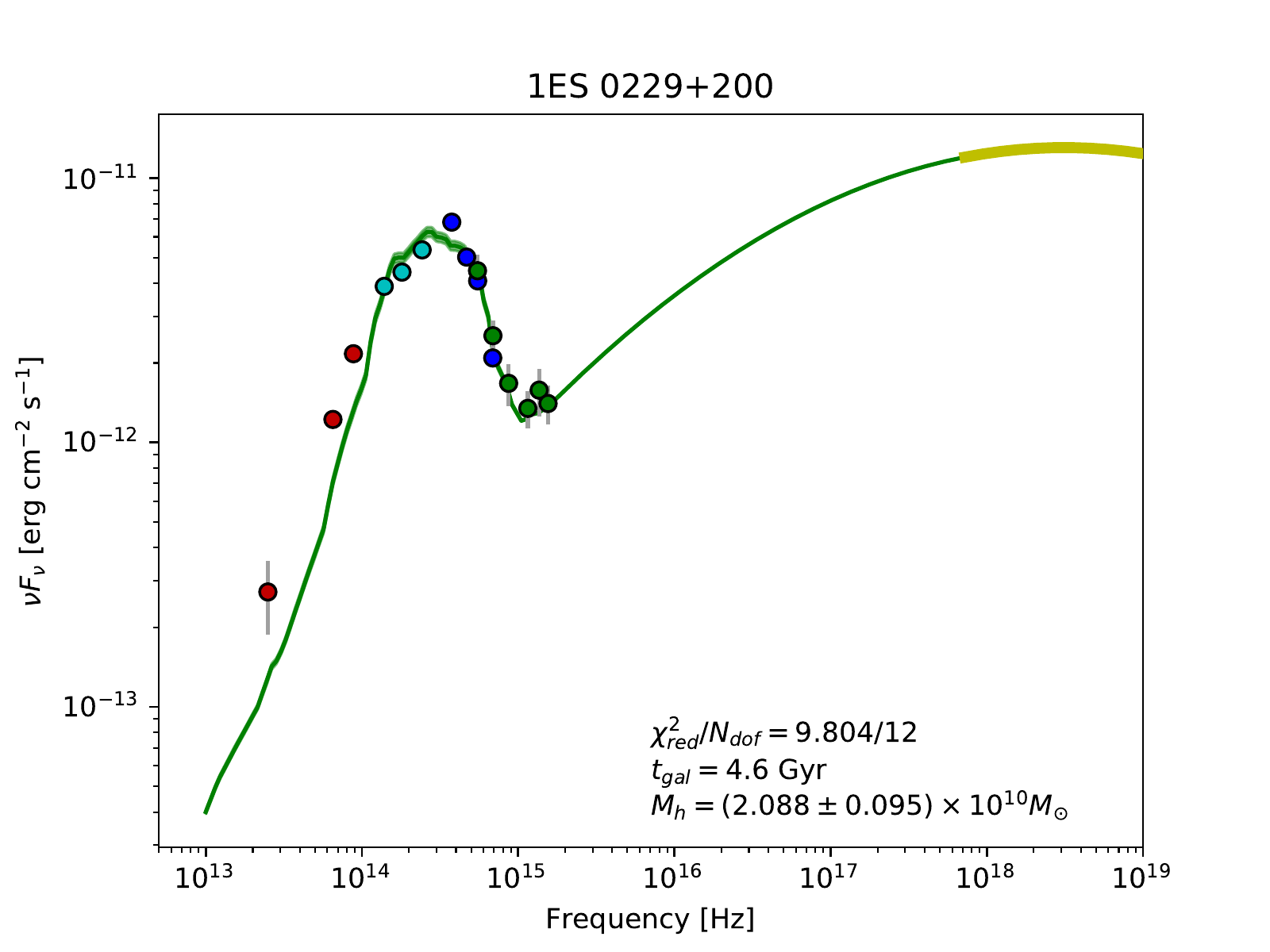}}
\centering{\includegraphics[width=0.3\textwidth]{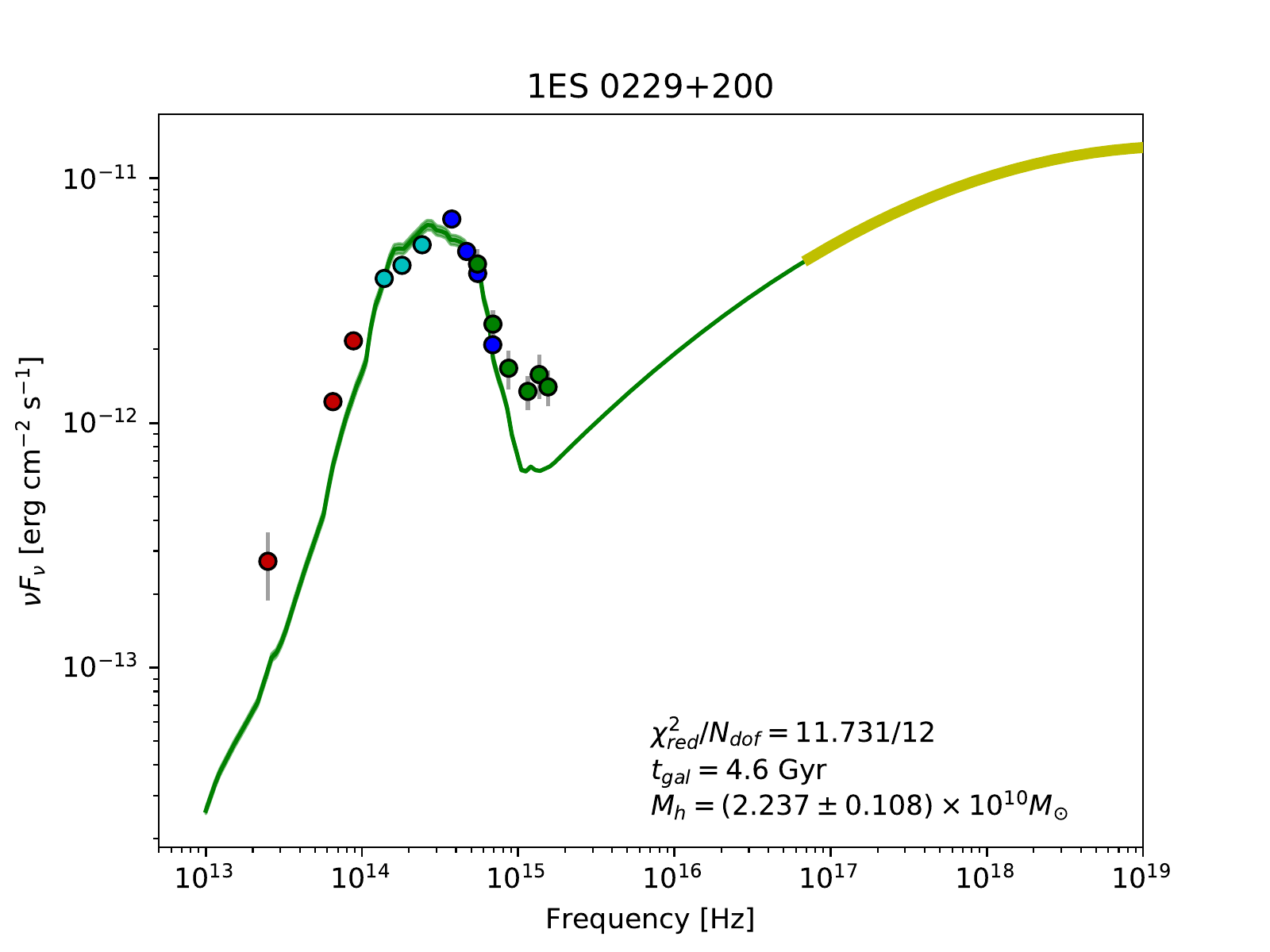}}
\centering{\includegraphics[width=0.3\textwidth]{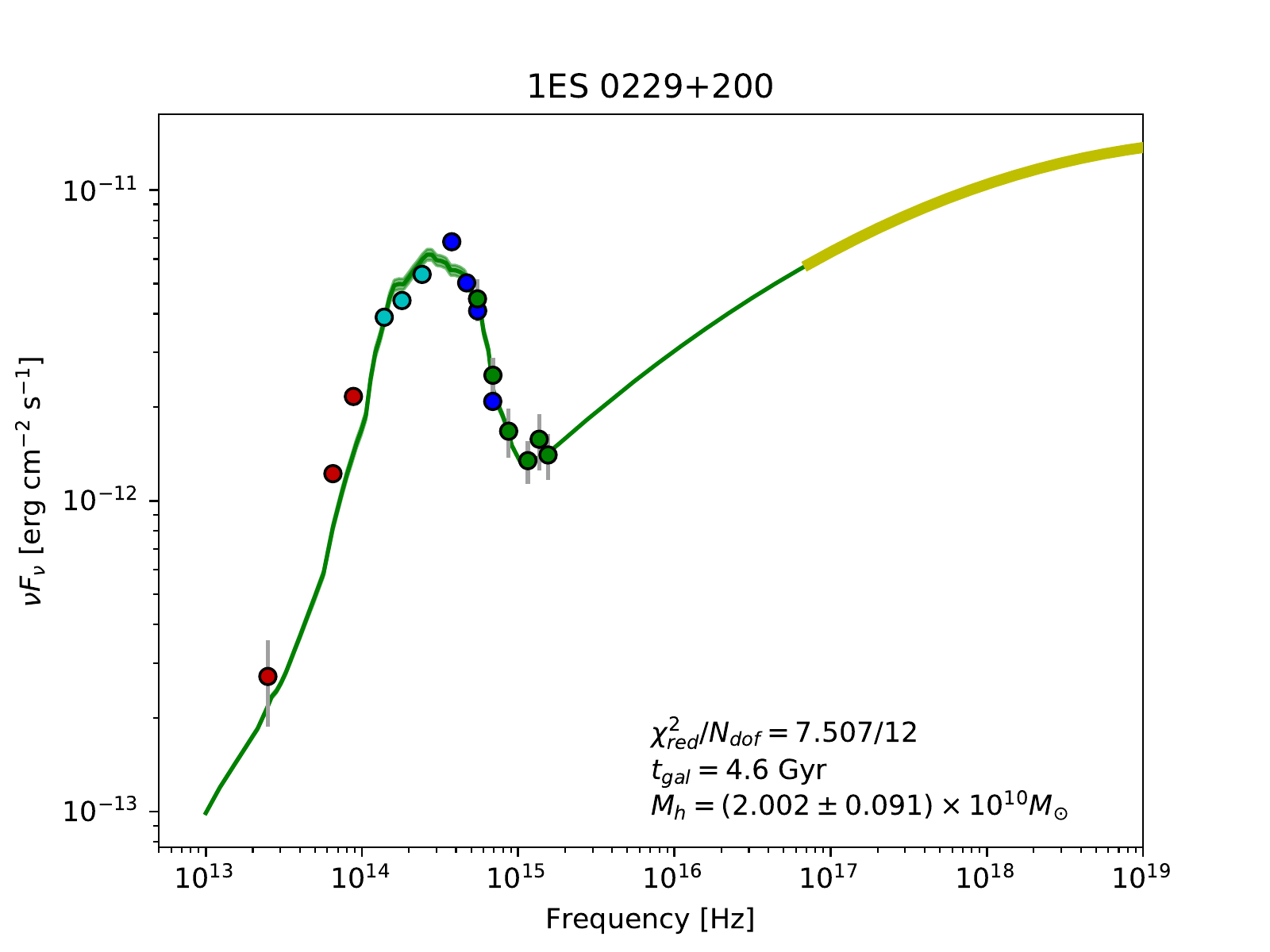}} \\
\centering{\includegraphics[width=0.3\textwidth]{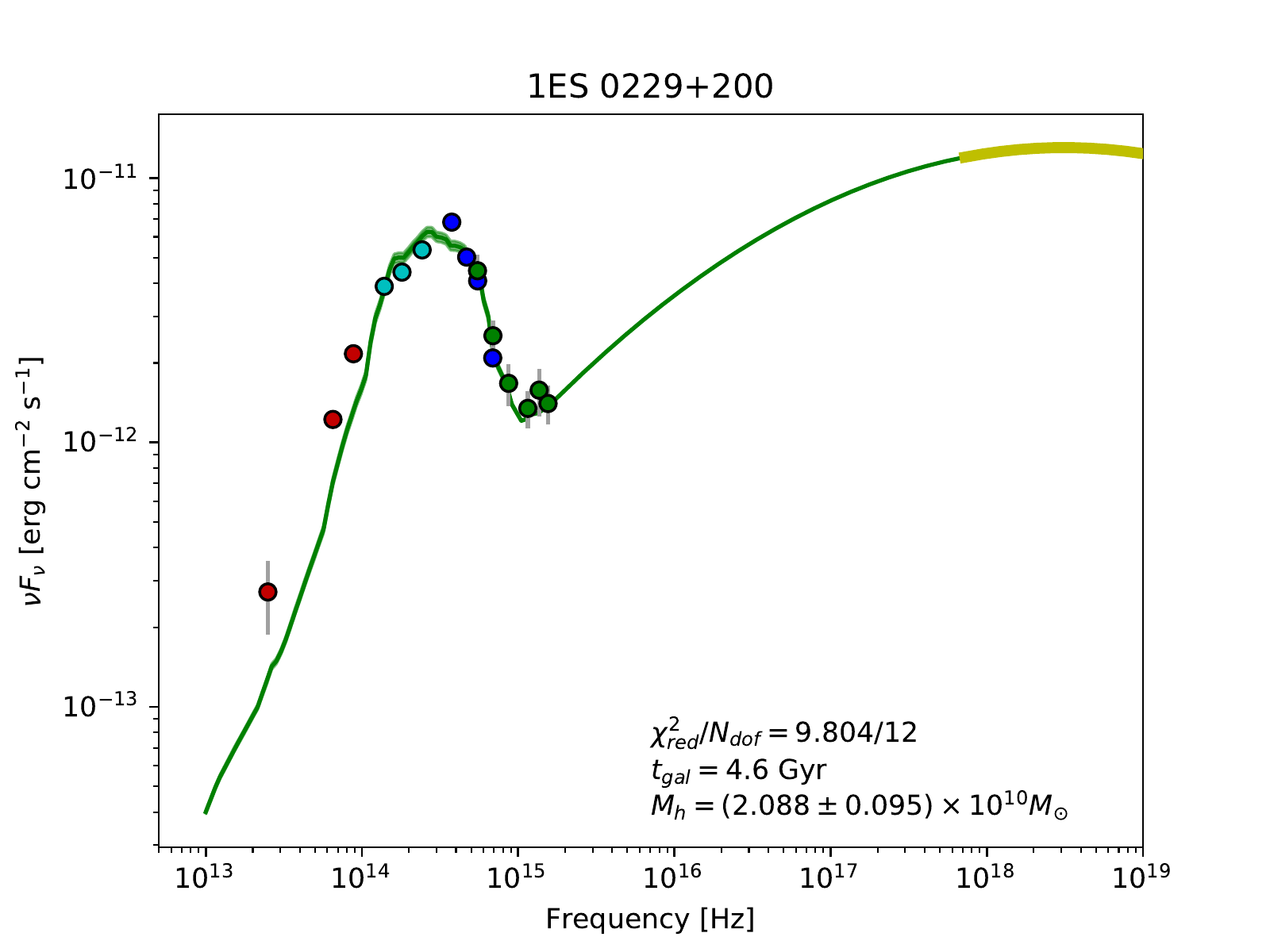}}
\centering{\includegraphics[width=0.3\textwidth]{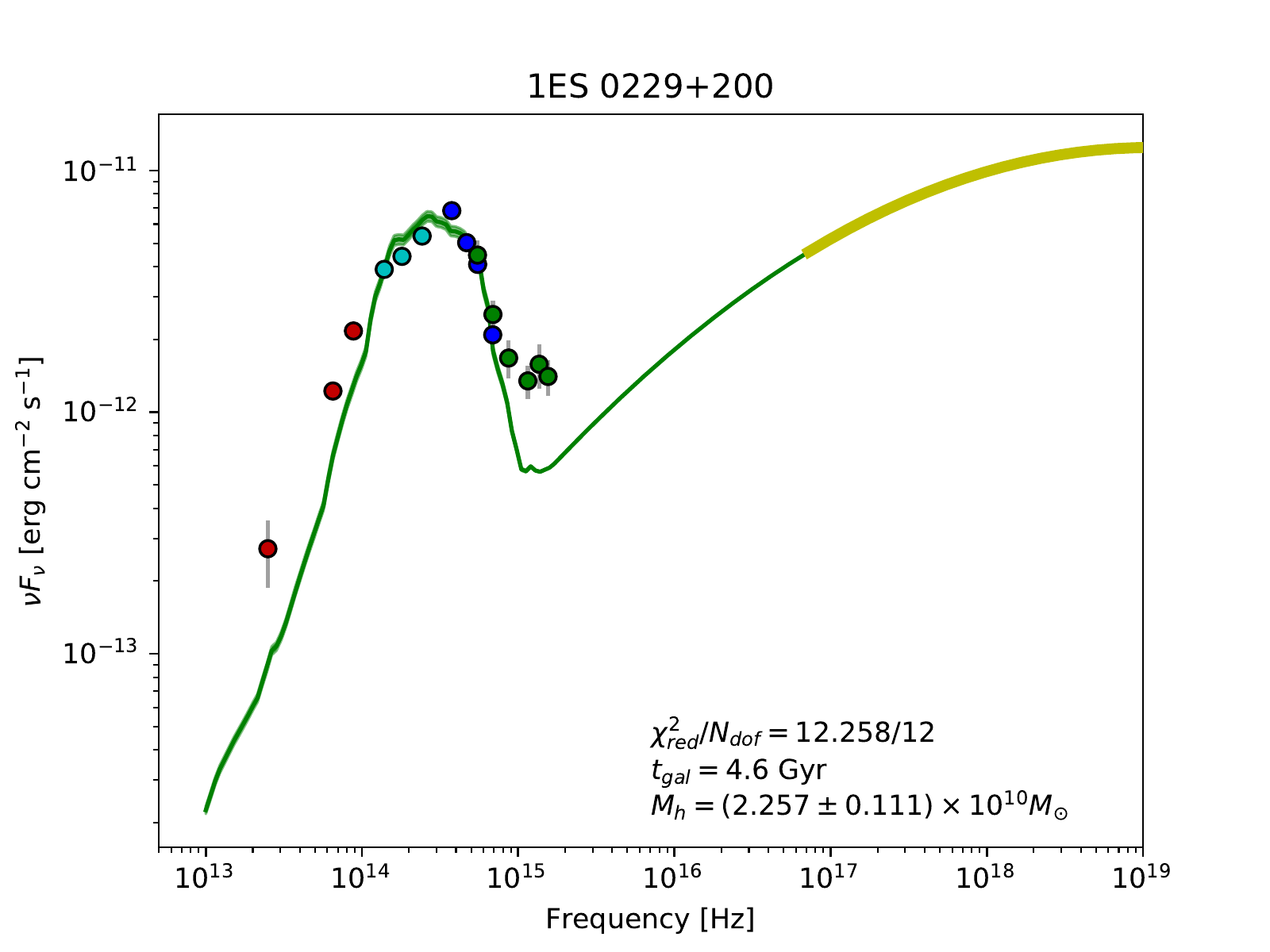}}
\centering{\includegraphics[width=0.3\textwidth]{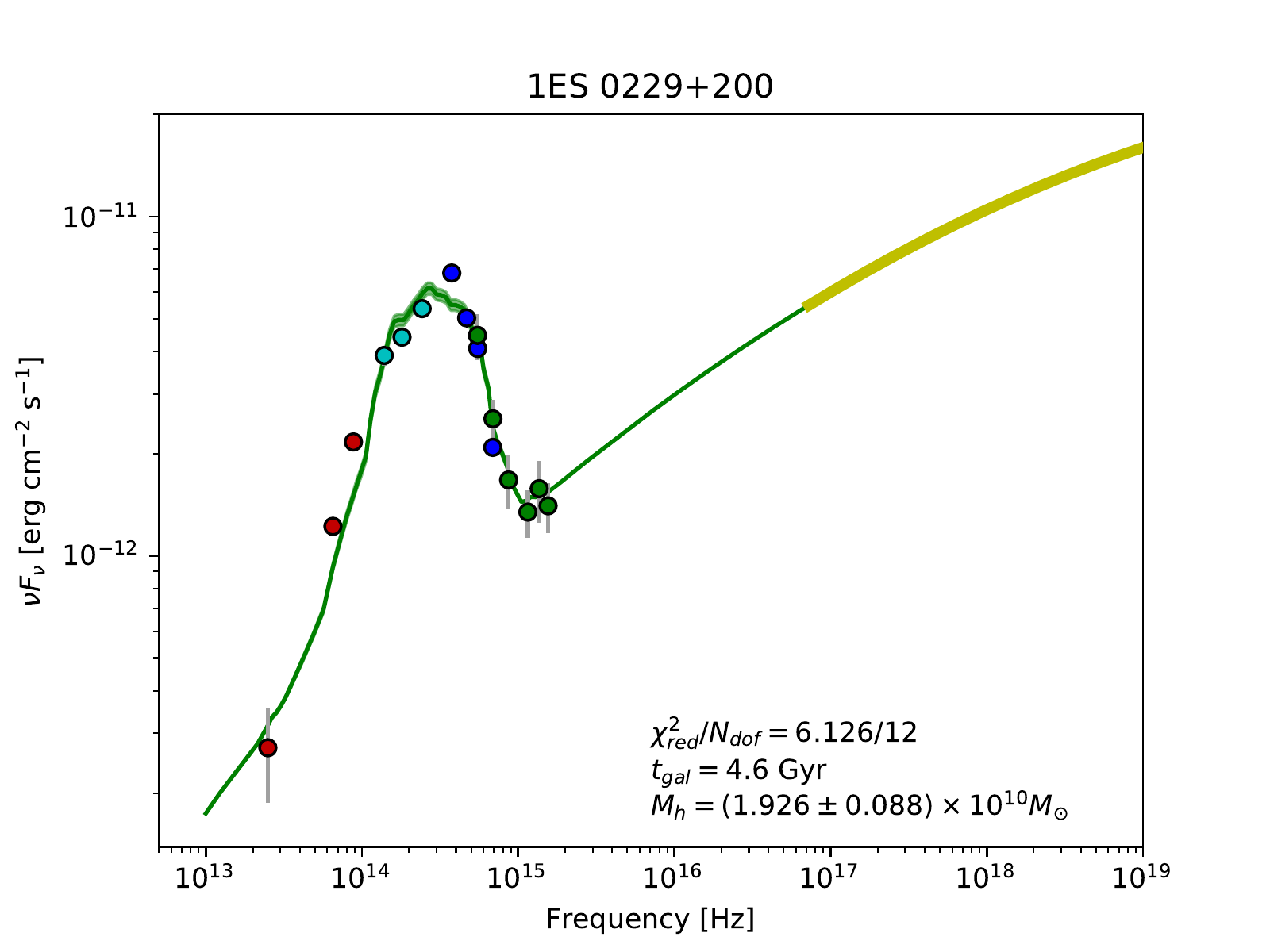}} \\

\caption[]{Spectral energy distribution of \one\ derived from the extrapolated \nus\ X-ray spectra.
First row - observations with the ObsID of nu60002047002; from left to right: the \lp\ fit to the \nus\ data only in the energy range of 3-79\,keV, the joint \xrt-\nus\  fit in the energy range of 0.3-79\,keV with the \lp\ model  and \nh\ value as provided by \cite{Kalberla2005}, the joint \xrt-\nus\  fit in the energy range of 0.3-79\,keV with the \lp\ model  and \nh\ value as provided by \cite{Willingale13}. 
Second row: same as first but for the ObsID of nu60002047004. 
Last row: same as first but for the ObsID of nu60002047006.  }
\label{figure:nu}
\end{figure*}

In order to test whether the \nus\ X-ray spectra of \one\ are better constrained with a single \po\ model or a \lp, all three \nus\ observations were fitted with  both models.
Table~\ref{table:results_0229} summarizes parametres of spectral fits and Figure~\ref{figure:nus_spec} shows three \nus\ spectra. 
We note here that in the case of the  \nus\ observations of \one\ X-ray spectrum is here constrained in the energy range of 3-50\,keV. 
Similar $\chi_{red}^2$ values for both \po\ and \lp\ fits indicate that the hard X-ray spectrum suggests that the spectra can be described with the same precision while using one of these models. 
However, using spectral parameters for the \po\ model of the \nus\ spectra, together with the host galaxy template, we are not able to reproduce the \itu\ SED of \one. 
The extrapolated fit was above the datapoints. 
Thus, we conclude that the proper description of the X-ray spectrum of \one\ in the energy range above 3\,keV can be done only with a curved model. 
As the \po\ model was ruled out, we used the logparbolic description of the X-ray spectra together with the host galaxy templates. 
The resulting SEDs are presented in the first column in Figure~\ref{figure:nu}. 
For all ObsID, none of the marginal UV excesses is visible.

Simultaneous \xrt-\nus\ observations of \one\ allow for joint spectral fitting in the energy range of 0.3-50\,keV. 
Figure~\ref{figure:nu} (the second and third column) presents broadband SEDs of \one\ being the result of \lp\ spectral model extrapolated together with the host galaxy template. 
Two different values of \nh\ are tested: the one provided by \cite{Kalberla2005} and by \cite{Willingale13}.
The UV excess is present in the SEDs obtained while using \nh\ by \cite{Kalberla2005}, and it is now visible in the case of SEDs obtained with \nh\ by e.g. \cite[][]{Willingale13}.

The analysis of \nus\ and \xrt-\nus\ data allows us concluding that while using  X-ray data above  10\,keV, we received as good information about synchrotron emission of \one\ as in the case of \xrt\ data only.

\subsubsection*{PKS\,0548-322}
\cite{Wierzcholska2016}  have shown that the longterm integrated X-ray spectrum of PKS\,0548-322 in the energy range of 0.3-10\,keV has been well described with logparabola model with Galactic column density value found by \cite{Kalberla2005}. 
The X-ray spectrum fit parameters provided by \cite{Wierzcholska2016} together with the galaxy template reproduce the optical and IR observations well, but significant excess in the UV band is observed.
In order to check the influence of the Galactic absorption on X-ray spectral parameters, the X-ray spectrum has been again fitted in the energy range of 2.0-10\,keV. 
The spectrum has been extrapolated together with the template of the host galaxy up to the \itu\ range. The resulting fit parameters are only slightly different from the 0.3-10\,keV fit, but they allow to reproduce broadband emission of    PKS\,0548-322 without significant UV excess visible. 
From this fit, the host galaxy of the blazar is described with a mass of $(4.7\pm0.3) \cdot 10^{9} M_{\astrosun}$. 
Fit parameters of the X-ray spectrum for both cases are collected in Table~\ref{table:results} and plots are demonstrated in Figure~\ref{figure:spectra_other}.
The studies performed have shown that additional absorption is needed to reproduce the SED of the blazar without significant UV excess.

PKS\,0548-322 was a frequent target of X-ray observations. 
The analysis reported by \cite{Costamante_0548} indicates the evidence of the intrinsic curvature of the X-ray spectrum of the blazar, even in the presence of an additional absorption. 
\cite{Perri_0548} also have reported a logparabola as a preferred model describing the X-ray spectrum of the blazar as seen with \xrt\ and Beppo-SAX.
\cite{hess_0548} have described the X-ray spectrum of PKS\,0548-322 in the energy range using a single \po\ and broken \po\ model without significant improvement while using the later model. However, the authors have mentioned a possible presence of an absorption feature near 0.7\,keV.

\subsubsection*{RX\,J1136+6737}
\cite{Wierzcholska2016} have found the power-law model as a good description of the X-ray spectrum in the energy range of 0.3-10\,keV. The authors have not  found any need for an additional absorption component in the spectral description. The preferred value of the Galactic column density has been the one provided by  \cite{Willingale13}.
Using extrapolated power-law fit together with a template of the host galaxy fits well the multiwavelength data without any excess in the UV range.  
A mass of host galaxy is estimated to have $(2.23 \pm 0.14)\cdot 10^{10}\,M_{\astrosun}$.
Fit parameters of the X-ray spectrum is presented in Table~\ref{table:results} and SED is shown in Figure~\ref{figure:spectra_other}.

\subsubsection*{1ES\,1741+196}
\cite{Wierzcholska2016} have shown that the X-ray spectrum of 1ES\,1741+196 is well described using a logparabola model with Galactic column density value provided by \cite{Willingale13}.
These spectral parameters together with the host galaxy template  were used and fitted to multiwavelength data collected for 1ES\,1741+196 to reproduce the low energy bump of the SED since significant UV excess is evident. 
The second fit to the data using the X-ray spectrum in the energy of 2.0-10\,keV reproduced data well without any excess visible in the UV range. 
According to this fit, the estimated value of the host galaxy mass is  $(1.37 \pm 0.08)\cdot 10^{10} \,M_{\astrosun}$.
The fit parameters of the X-ray spectrum is presented in Table~\ref{table:results} and SED is shown in Figure~\ref{figure:spectra_other}.
The studies performed have shown that additional absorption is needed to reproduce the SED of the blazar without significant UV excess. 

\cite{MAGIC_1741} have found that the X-ray spectrum of 1ES\,1741+196 is well described with a single \po\ without any indication of absorption excess. 
Contrary to these studies, \cite{VERITAS_1741} have concluded analysis of \xrt\ spectra of 1ES\,1741+196 that curvature in the spectrum is needed in order to explain the spectral shape in this range.

\subsubsection*{1ES\,2344+514}
\begin{figure*}
\centering{\includegraphics[width=0.4\textwidth]{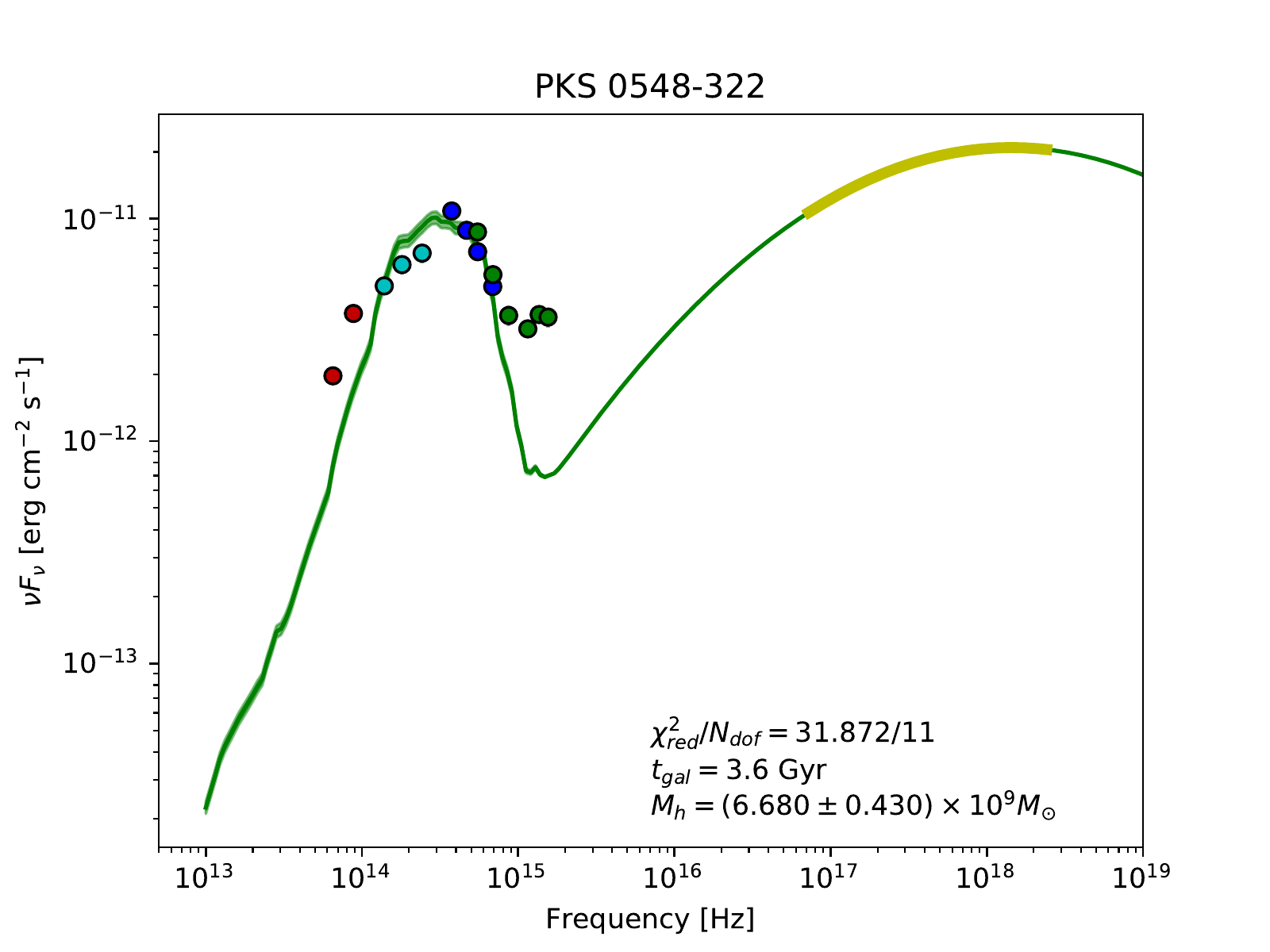}}
\centering{\includegraphics[width=0.4\textwidth]{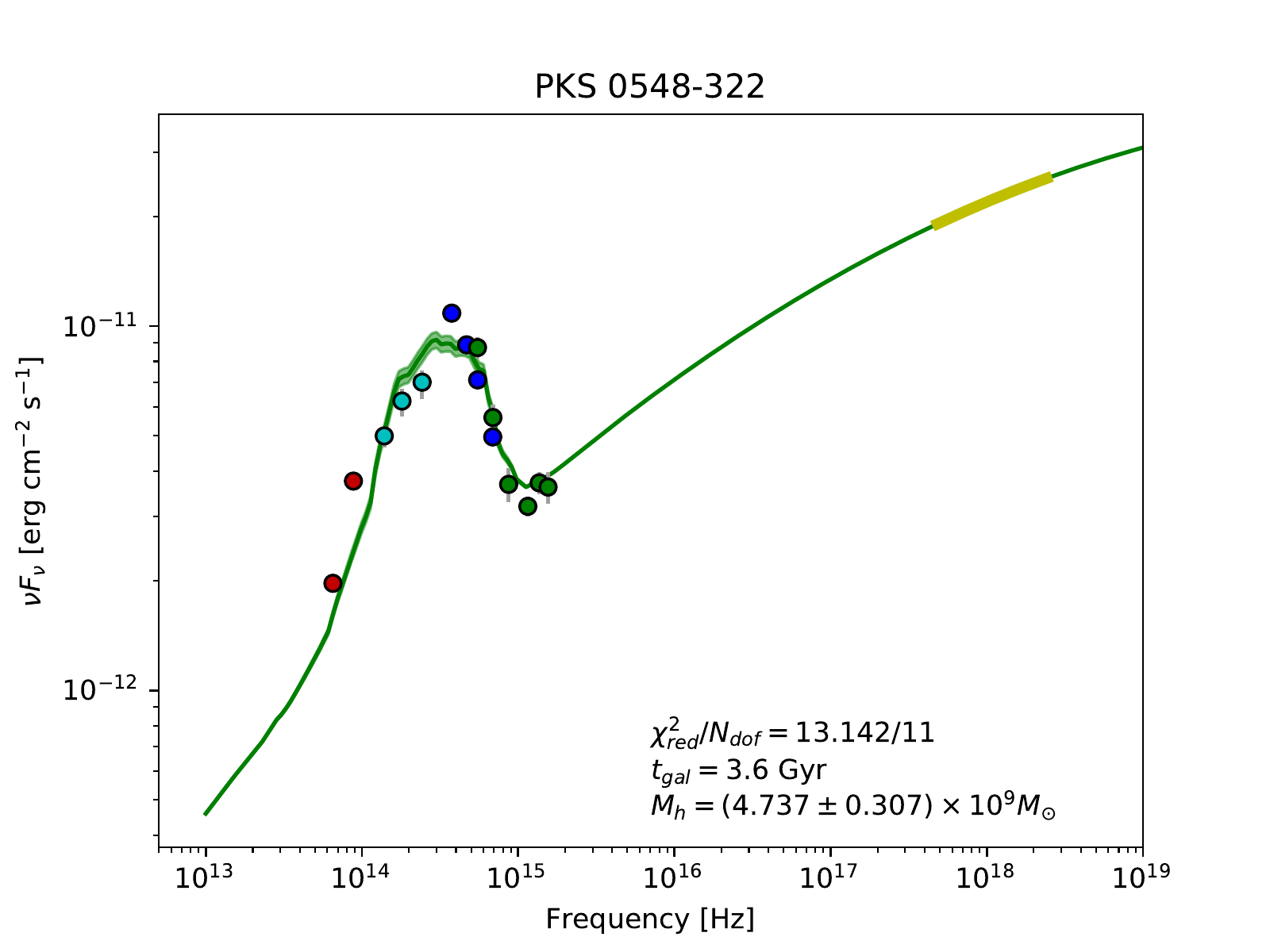}}
\centering{\includegraphics[width=0.4\textwidth]{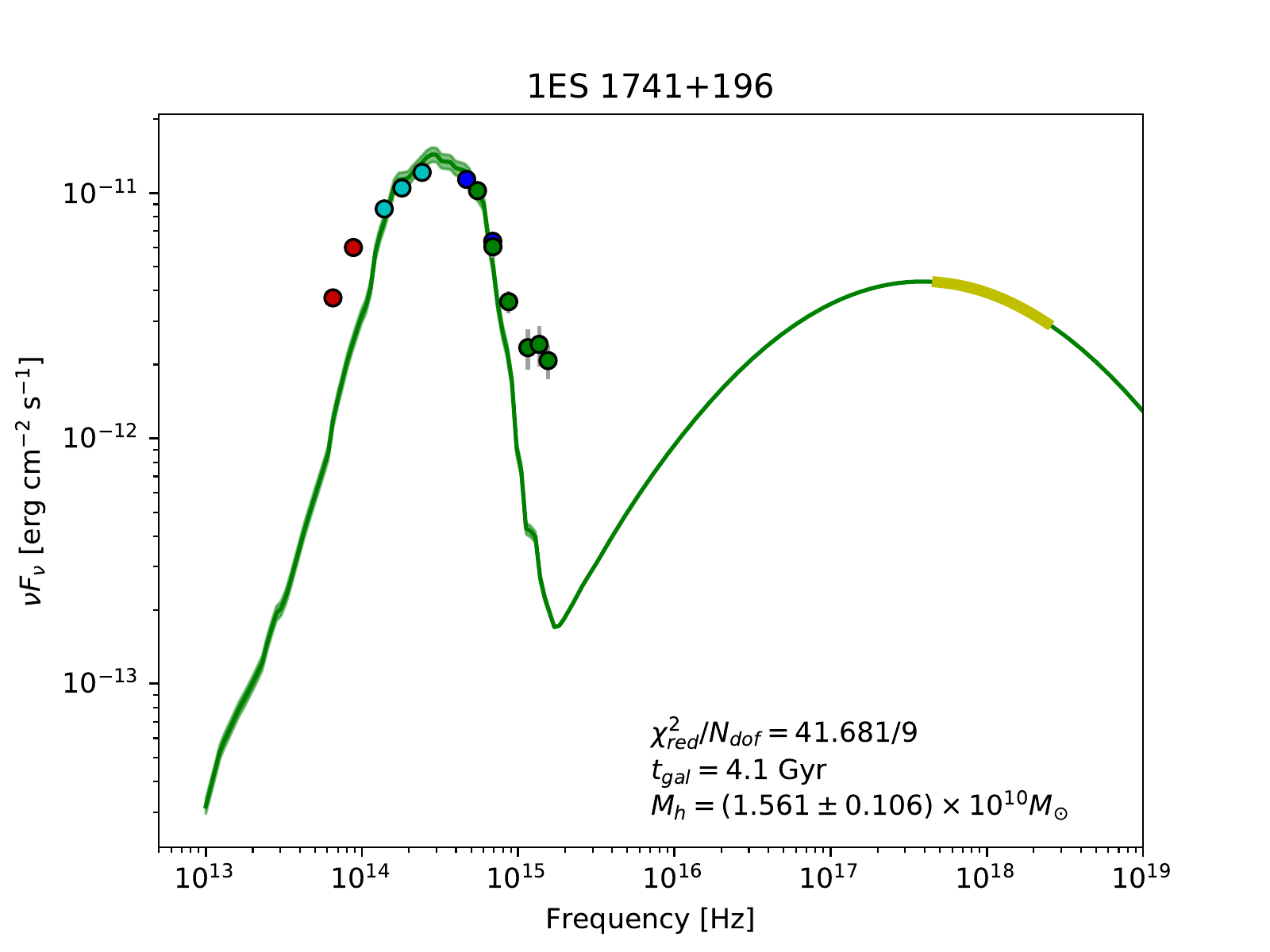}}
\centering{\includegraphics[width=0.4\textwidth]{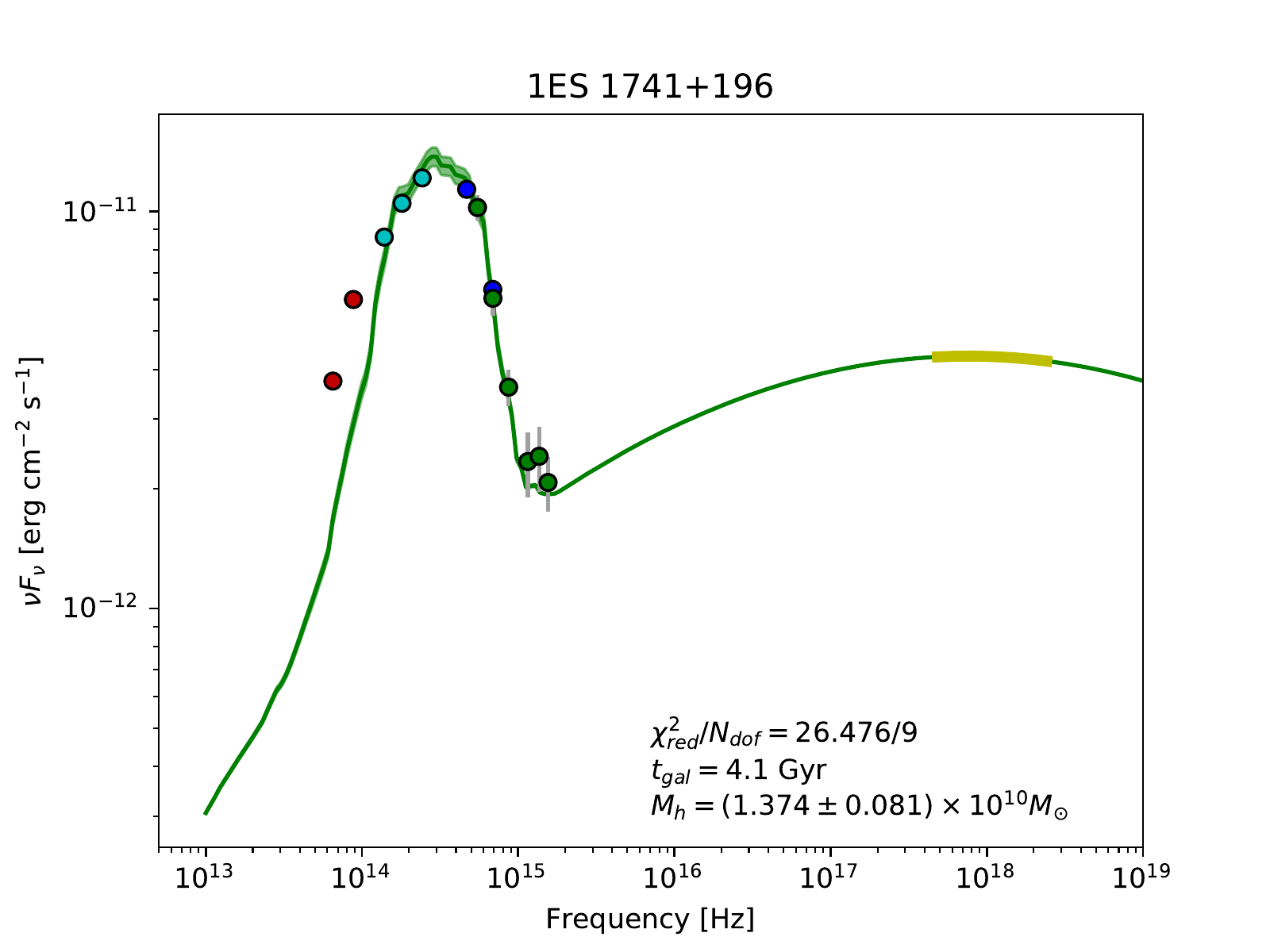}}
\centering{\includegraphics[width=0.4\textwidth]{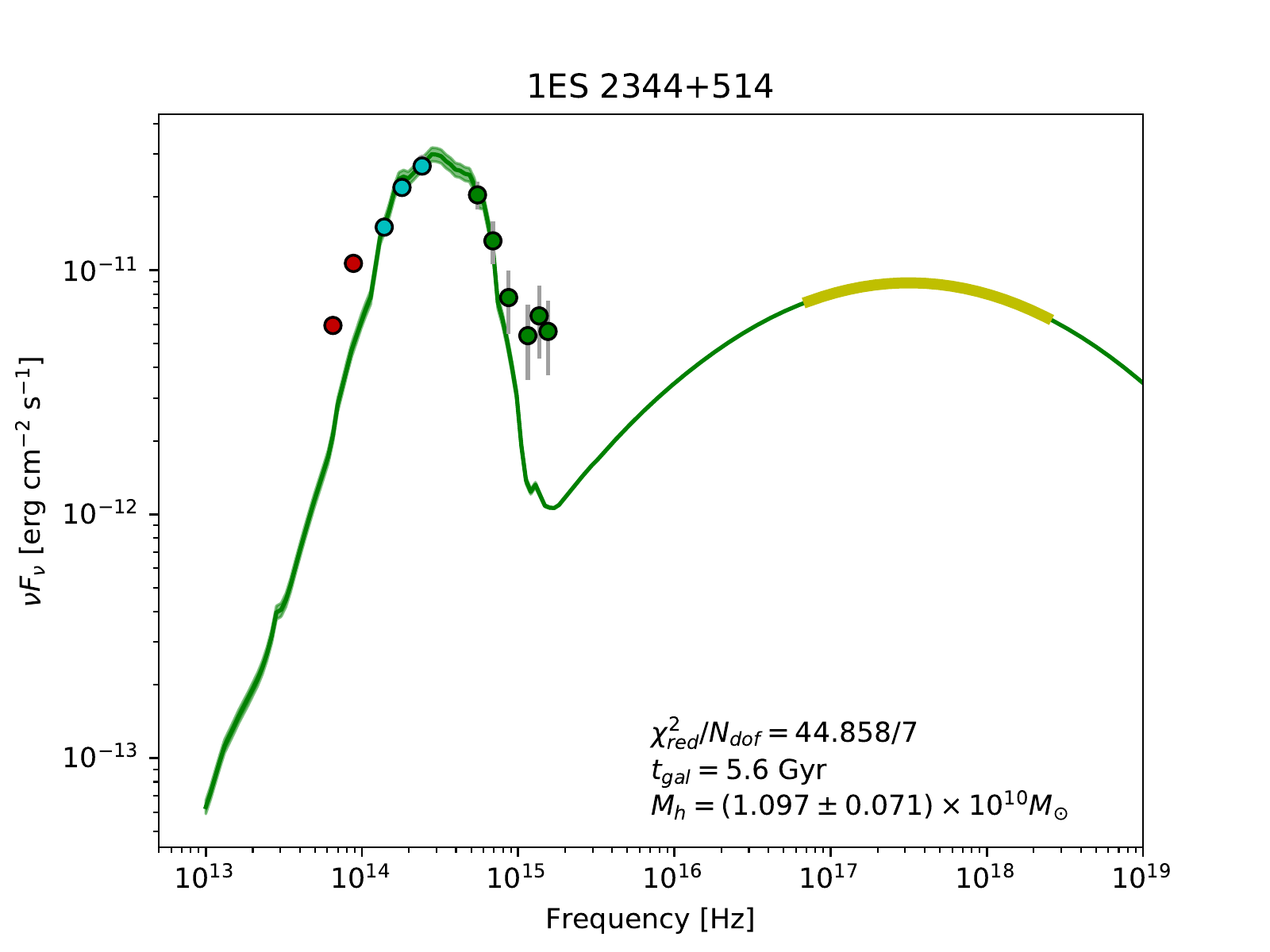}}
\centering{\includegraphics[width=0.4\textwidth]{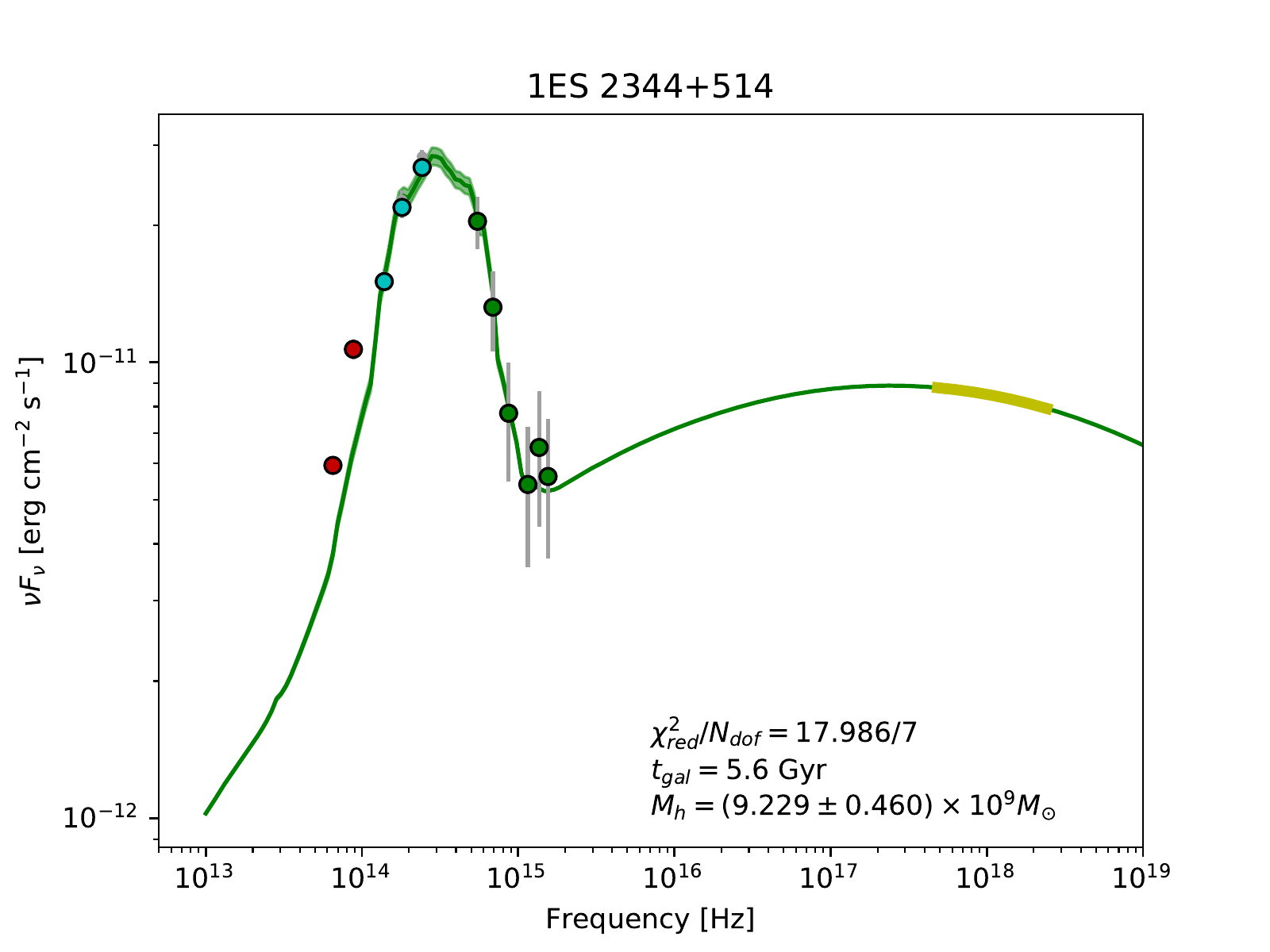}}
\centering{\includegraphics[width=0.4\textwidth]{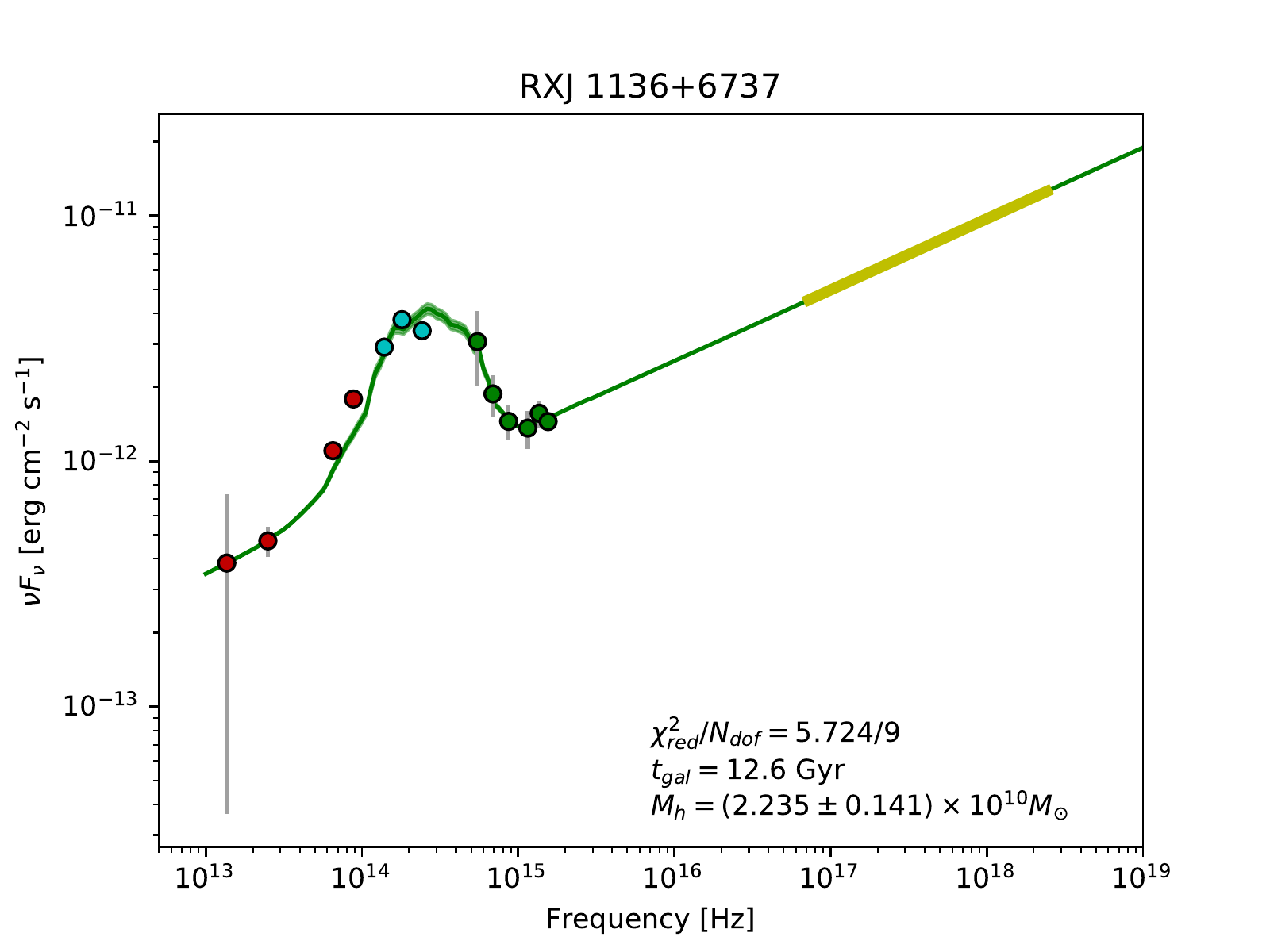}}
\caption[]{From top to bottom: Broadband SED of PKS\,0548-322. Left: modelling with X-ray spectrum fitted with the logparabola model in the energy range of 0.3-10\,keV; right: same as left but X-ray spectrum fitted in the energy range of 2.0-10\,keV.;
Broadband SED of 1ES\,1741+196. Left: modelling with X-ray spectrum fitted with the logparabola model in the energy range of 0.3-10\,keV; right: same as left but X-ray spectrum fitted in the energy range of 2.0-10\,keV.;
Broadband SED of 1ES\,2344+514. Left: modelling with X-ray spectrum fitted with the logparabola model in the energy range of 0.3-10\,keV; right: same as left but X-ray spectrum fitted in the energy range of 2.0-10\,keV.
Broadband SED of RXJ\,1136+6737: modelling with X-ray spectrum fitted with the power-law model in the energy range of 0.3-10\,keV. 
Points colour coding same as in Fig.\ref{figure:spectra_global}.}
\label{figure:spectra_other}
\end{figure*}

In the studies of longterm X-ray spectra of blazars \cite{Wierzcholska2016} have shown that the spectrum of 1ES\,2344+514 is well described with the logparabola model with Galactic column density taken from the survey by \cite{Kalberla2005}.
This X-ray spectral fit extrapolated to the \itu\ range, together with a host galaxy template, results in large UV excess. 
Fit to the same X-ray data, but in the energy range of 2.0-10\,keV, together with the host galaxy, gives appropriate fit without UV excess. 
For the second fit estimated value of  host galaxy mass is $(19.2\pm00.5)\cdot 10^{10} \,M_{\astrosun}$.
Results of fits for both cases can be found in Fig.~\ref{figure:spectra_other}, and parameters of all fits are presented in Table~\ref{table:results}. 
The studies performed have shown that additional absorption is needed to reproduce the SED of the blazar without significant UV excess. 

No need for a curved model describing spectrum or additional absorbing component has been reported by \cite{MAGIC_2344}.
\cite{Kapanadze_2344} also have  prefered \po\ model to describe the X-ray spectra of this blazar. 
 
  We note here also, that for the case of 1ES\,2344+514 due to the low Galactic latitude of the blazar the \nh\ values taken from different catalogues are affected by large uncertainties.

\subsection{\nh\  calculation}
 The spectral parameters of the fit in the energy range of  2-10\,keV were used as frozen for the fit to data in the energy range of 0.3-10\,keV. 
The only free fit parameter here was \nh, and thus, we were able to calculate \nh\ values for all  targets studied.

 \begin{itemize}
 \item
 For 1ES\,0229+200 the calculated value of N$_H^{tot}$ is (13.00$\pm$0.37)$\cdot$10$^{20}$cm$^{-2}$. This value is larger than the one from the LAB survey (8.06$\cdot$10$^{20}$cm$^{-2}$) and consistent with the one provided by \cite{Willingale13} (10.80$\cdot$10$^{20}$cm$^{-2}$). 
 The value is also higher than free \nh = (10.6-10.8)$\cdot$10$^{20}$cm$^{-2}$ calculeted by \cite{Kaufmann2011}.
 \item
 For PKS\,0548-322 the calculated value of N$_H^{tot}$ is (3.10$\pm$0.10)$\cdot$10$^{20}$cm$^{-2}$. This value is larger than the one from the LAB survey (2.58$\cdot$10$^{20}$cm$^{-2}$)  and consistent with the one  provided by \cite{Willingale13} (2.87$\cdot$10$^{20}$cm$^{-2}$).
 However, \cite{Sambruna_1998} have found significatly higher value of \nh\ of 1.03$\cdot$10$^{21}$cm$^{-2}$ that is needed to explain X-ray properties of PKS\,0548-322. 
  \item
 For 1ES\,1741+196 the calculated value of N$_H^{tot}$ is (10.47$\pm$0.41)$\cdot$10$^{20}$cm$^{-2}$. This value is larger than the one from the LAB survey (7.32$\cdot$10$^{20}$cm$^{-2}$)  and consistent with the one  provided by \cite{Willingale13} (9.58$\cdot$10$^{20}$cm$^{-2}$).
   \item
 For 1ES\,2344+514 calculated value of N$_H^{tot}$ is (17.34$\pm$0.15)$\cdot$10$^{20}$cm$^{-2}$. This value is larger than the one from the LAB survey (15.10$\cdot$10$^{20}$cm$^{-2}$), however it is smaller than  the one provided by \cite{Willingale13} (24.60$\cdot$10$^{20}$cm$^{-2}$).
\end{itemize}

We note here  that the reddening $E(B-V)$ and the atomic hydrogen column density (\nh) are related.
 According to recent work, \cite{Liszt2014} has described this dependence by the following formula \nh = $8.3 \cdot$ 10$^{21}E(B-V)$.
The formula gives \nh\ values of 9.7$\cdot$ 10$^{20}$, 2.5$\cdot$ 10$^{20}$, 0.6$\cdot$ 10$^{20}$, 6.3$\cdot$ 10$^{20}$, and 15$\cdot$ 10$^{20}$ for  1ES\,0229+200, PKS\,0548-322, RX\,J1136+6737, 1ES\,1741+196, 1ES\,2344+514, respectively. 
The quoted values are consistent with the ones from the LAB survey and do not indicate a need for an additional absorption.
All values of \nh\ values, including the catalogue ones, are collected in Table\,\ref{table:nhvalues}.

\begin{table*} 
\centering 
\begin{tabular}{c|c|c|c|c|c|c|c|c|c}
\hline
\hline
 Source &  Model  & Energy range & N & $\gamma$ or $\alpha$ & $\beta$  & N$_H$  & $\chi_{red}^2$(n$_{d.o.f}$) &  M & ID  \\
(1) &  (2)& (3) & (4) & (5) & (6)  & (7) & (8) & (9) & (10)   \\
\hline
1ES\,0229+200 & power-law    & 0.3-10 & 3.70$\pm$0.05 & 1.59$\pm$0.02 & - & 8.06 & 1.38(266)   &2.15$\pm$0.10 &  A\\
1ES\,0229+200 & logparabola  & 0.3-10 & 3.81$\pm$0.06 & 1.41$\pm$0.03 & 0.37$\pm$0.05 & 8.06 & 1.16(265)   & 2.36$\pm$0.13 &  B \\
1ES\,0229+200 & power-law    & 2.0-10 & 4.01$\pm$0.06 & 1.67$\pm$0.02 & - & 8.06 & 0.98(114)   & 1.97$\pm$0.09 & C \\
1ES\,0229+200 & power-law    & 0.3-10 & 4.21$\pm$0.06 & 1.70$\pm$0.02 & - & 11.8 & 1.20(266)   & 1.87$\pm$0.09 & D  \\
\hline
PKS\,0548-322 & logparabola  & 0.3-10 & 10.28$\pm$0.06 & 1.73$\pm$0.01 & 0.17$\pm$0.02 & 2.58 & 1.09(750)    &  0.67$\pm$0.04 & E  \\
PKS\,0548-322 & logparabola  & 2.0-10 & 10.29$\pm$0.06 & 1.78$\pm$0.01 & 0.04$\pm$0.01 & 2.58 &   1.10(584)   &0.47$\pm$0.03 & F \\
\hline
RXJ\,1136+6737 & power-law  & 0.3-10 & 4.02$\pm$0.04 & 1.73$\pm$0.01 & - & 1.06 &  1.11(468)   & 2.08$\pm$0.13 & G \\
\hline
1ES\,1741+196 & logparabola  & 0.3-10 & 2.55$\pm$0.05  & 1.89$\pm$0.03 & 0.26$\pm$0.01 & 9.58 &  1.05(350)   & 1.56$\pm$0.10 & H  \\
1ES\,1741+196  & logparabola  & 2.0-10 & 2.62$\pm$0.05 & 1.95$\pm$0.02 & 0.05$\pm$0.03 & 9.58 &  0.98(186)   & 1.39$\pm$0.08 & I \\
\hline
1ES\,2344+514 & logparabola  & 0.3-10 & 5.51$\pm$0.03 & 1.96$\pm$0.01 & 0.18$\pm$0.02 & 15.1 &   1.03(707)   & 1.10$\pm$0.07 & J \\
1ES\,2344+514 & logparabola  & 2.0-10 & 5.55$\pm$0.04 & 2.00$\pm$0.01 & 0.05$\pm$0.03 & 15.1 &   1.00(539)   &0.92$\pm$0.05  & K \\

\hline
\end{tabular}
\caption[]{Fit parameters and results for the sample:  (1) name of the source;  (2) model used to describe X-ray observations; (3) energy range of the X-ray spectrum;  (4) normalization given in 10$^{-3}$ph\,cm$^{-2}$\,s$^{-1}$\,keV$^{-1}$; (5) photon index: $\gamma$ in case of power-law model and $\alpha$ in the case of logparabola one; (6) curvature coefficient $\beta$ for logparabola model; (7) Galactic absorption value, given in 10$^{20}$ cm$^2$; (8) $\chi_{red}^2$ for a fit of X-ray data with a number of degree of freedom; (9)   Estimated mass of the host galaxy given in 10$^{10}$M$_{\astrosun}$; (10) ID of the fit.    }
  \label{table:results}
\end{table*}


\begin{table} 
\centering 
\begin{tabular}{c|c|c|c|c|c}
\hline
\hline
 Source &  LAB  & DL & Will & N$_{E(B-V)}$  &  this work   \\
(1) &  (2)& (3) & (4) & (5) & (6) \\
\hline
1ES\,0229+200 & 8.06  & 9.20  & 11.80 & 9.7 &13.00$\pm$0.37 \\
PKS\,0548-322 & 2.58  & 2.18  & 2.87 & 2.5 &3.01$\pm$0.10 \\
RXJ\,1136+6737 & 1.03 & 1.33  & 1.06& 0.6& -- \\
1ES\,1741+196 & 7.32 & 6.82  & 9.58 & 6.3 & 10.47$\pm$0.41 \\
1ES\,2344+514 & 15.10 &  16.80 & 24.60 & 15 &17.34$\pm$0.15 \\

\hline
\end{tabular}
\caption[]{Summary of Galactic column densities values. The following columns presents: (1) source name; (2)-(4) \nh\ as provided by \cite{Kalberla2005}, \cite{Dickey1990}, \cite{Willingale13}, respectively; (6) \nh\ value calcuted as a result of this work. 
All values are given in 10$^{20}$cm$^{-2}$. }
  \label{table:nhvalues}
\end{table}

\noindent

\begin{table*} 
\centering 
\begin{tabular}{c|c|c|c|c|c}
\hline
\hline
 NuSTAR obsId & Time   & NuSTAR exposure [s] & \xrt\ obsIds & Time & Exposure [s]\\
\hline
60002047002 &  2013-10-02 00:06:07 & 16258 & 00080245001  &  2013-10-01 23:36:59 & 612 \\
            &                      &       &  00080245003  & 	2013-10-02 00:45:59  & 3209 \\

60002047004 &  2013-10-05 23:31:07 & 20291  &  00080245004 &  2013-10-06 00:51:59 &  5277\\

60002047006 & 2013-10-10 23:11:07   &  18023 &  00080245005 & 2013-10-10 23:47:58 &581  \\
            &                      &       &  00080245006  & 2013-10-11 01:29:59  & 5417 \\
\hline
\end{tabular}
\caption[]{Summary of \nus\ and \xrt\ joint  observations of \one.}
  \label{table:nustar}
\end{table*}

\begin{table*}

\caption{Fits and results for \one.    (1) model used to describe X-ray observations; (2) energy range of the X-ray spectrum;  (3) normalization given in 10$^{-3}$ph\,cm$^{-2}$\,s$^{-1}$\,keV$^{-1}$; (4) photon index: $\gamma$ in case of power-law model and $\alpha$ in the case of logparabola one; (5) curvature coefficient $\beta$ for logparabola model; (6) Galactic absorption value, given in 10$^{20}$ cm$^2$; (7) $\chi_{red}^2$ for a fit of X-ray data with a number of degree of freedom (8)  Additional information i.e. \nus\ observation ID.}
\label{table:results_0229}

\centering

\begin{tabular}{c|c|c|c|c|c|c|c}
\hline
\hline
  Model  & Energy range & N & $\gamma$ or $\alpha$ & $\beta$  & N$_H$  & $\chi_{red}^2$(n$_{d.o.f}$) &  NuSTAR obsId  \\
(1) &  (2)& (3) & (4) & (5) & (6)  & (7) & (8)    \\
\hline

power-law    &  3-79 &   6.67$\pm$0.20   &   1.96$\pm$0.04   &  -  & 8.06 &   1.019(86)      & nu60002047002\\
power-law    &  3-79 &   7.46$\pm$0.50   &   1.97$\pm$0.04   &  -  & 8.06 &   1.269(77)      & nu60002047004\\
power-law    &  3-79 &   7.60$\pm$0.50   &   1.97$\pm$0.04   &  -  & 8.06 &   1.040(72)      & nu60002047006\\
\hline
logparabola   &  3-79 &   5.53$\pm$0.13   &   1.78$\pm$0.07   &  0.10$\pm$0.02  & 8.06 &   1.012(85)      & nu60002047002\\
logparabola   &  3-79 &   6.35$\pm$0.15   &   1.80$\pm$0.20   &  0.10$\pm$0.03  & 8.06 &   1.280(76)      & nu60002047004\\
logparabola   &  3-79 &   6.35$\pm$0.20   &   1.80$\pm$0.15   &  0.10$\pm$0.02  & 8.06 &   1.058(71)      & nu60002047006\\
\hline

logparabola   &  0.3-79 &    5.30$\pm$0.10    &    1.70$\pm$0.10   &   0.16$\pm$0.02  & 8.06 &    1.015(130)      & nu60002047002\\
logparabola   &  0.3-79 &    4.41$\pm$0.09    &    1.70$\pm$0.11   &   0.09$\pm$0.03  & 8.06 &    1.224(154)      & nu60002047004\\
logparabola   &  0.3-79 &    4.36$\pm$0.08    &    1.70$\pm$0.11   &   0.08$\pm$0.01  & 8.06 &    1.331(175)      & nu60002047006\\
\hline
logparabola   &  0.3-79 &    6.15$\pm$0.03    &    1.80$\pm$0.07   &   0.10$\pm$0.01  & 11.8 &    1.028(130)      & nu60002047002\\
logparabola   &  0.3-79 &    4.88$\pm$0.10    &    1.77$\pm$0.11   &   0.06$\pm$0.01  & 11.8 &    1.260(154)      & nu60002047004\\
logparabola   &  0.3-79 &    4.70$\pm$0.09    &    1.75$\pm$0.11   &   0.04$\pm$0.01  & 11.8 &    1.044(175)      & nu60002047006\\

\hline

\end{tabular}
\end{table*}

\noindent

\subsection{Constrains on the model parameters}
The model used to describe the infrared to X-ray emission of the blazars is characterized by: host galaxy, curvature seen in the X-ray spectrum, total absorption (which includes both Galactic component and possibly additional ones), and ultraviolet excess. 
The host galaxy, understood as the entire component described with the template by \cite{Silvathesis},  is well constrained with the model used. Still, the remaining three parameters cannot be disentangled using only information from the X-ray spectral fitting. 
The limitations on three entangled parameters are the following:

\begin{itemize}
 \item \textbf{$\beta$}: 
 For all of the blazars studied, the convex curvature of the X-ray spectrum is assumed. This implies that negative values of $\beta$ are not accepted. 
 Furthermore, \cite{Wierzcholska2016} have demonstrated that in the case of a fixed value of absorption (e.g., from the LAB survey), X-ray spectra can be characterized with $\beta$ from 0.1 up to 0.9. 
Albeit, in the case of the free value of \nh, the value of $\beta$ ranges from 0.04 up to 0.4. 
In our studies, spectral curvatures of extreme HBL blazars range between 0.05 and 0.37, which is consistent with the work by \cite{Wierzcholska2016} and also with other studies focusing on spectral curvatures seen in the X-ray range \citep[][]{Massaro2004, Tramacere2007}.
 \item \textbf{\nh}: According to the LAB survey, values of the Galactic absorption range between 10$^{20}$-10$^{22}$.
For the blazars in the sample, \nh\ values do not exceed 10$^{21}$, and additional value required to explain UV excess is of about 60$\%$ higher than the survey ones, which is still in the range of reasonable value. 
 \item \textbf{UV excess}: The value of UV excess cannot be negative. 
\end{itemize}

To check the impact of a UV excess  on other parameters, for each source, we performed a study on parameters variability in the following steps:
\begin{itemize}
 \item  Range of possible \nh\ values is selected. The starting value is the lowest one from the surveys.
 \item  Spectral fitting of the X-ray spectrum in the energy range of 0.3-10\,keV with the frozen value of \nh\ is performed. 
 \item  Fitting to the host galaxy template as described in \ref{fitting}.
 \item  Calculation of the UV excess as a distance of the model to UVW2 datapoint.
\end{itemize}

\begin{figure*}
 \centering{\includegraphics[width=0.4\textwidth]{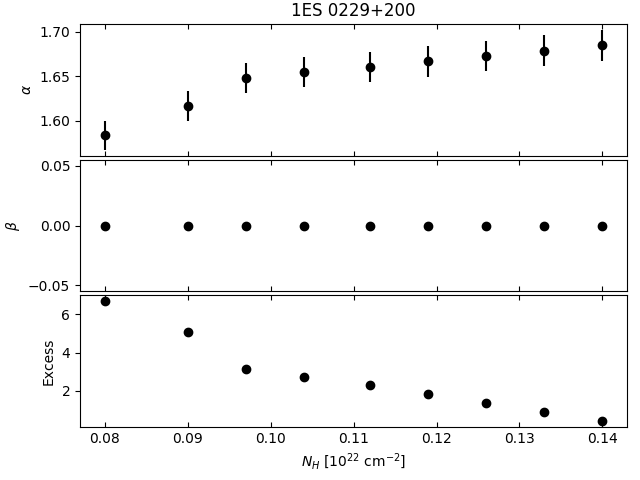}}
  \centering{\includegraphics[width=0.4\textwidth]{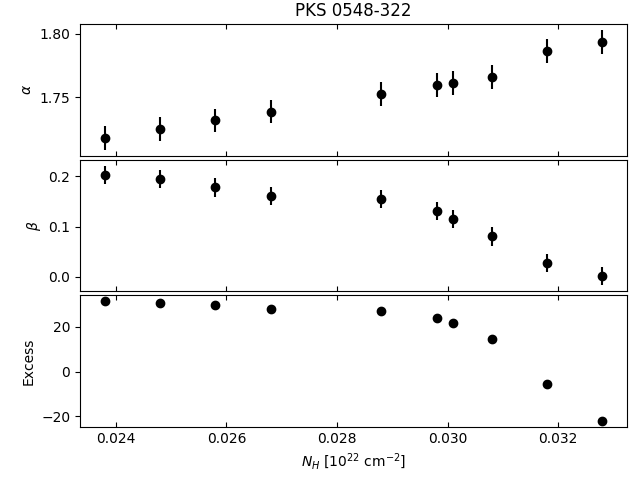}} \\
   \centering{\includegraphics[width=0.4\textwidth]{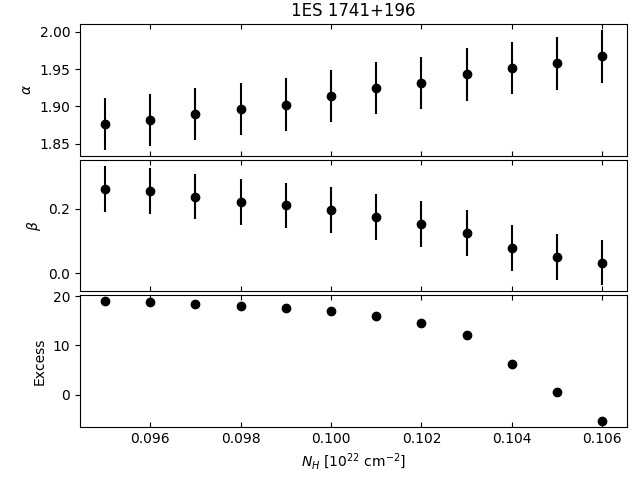}}
  \centering{\includegraphics[width=0.4\textwidth]{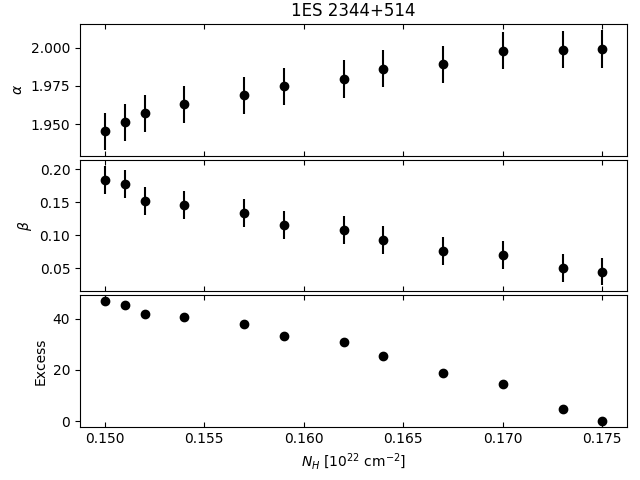}} \\
\caption[]{Comparison of model parameters: $\alpha$ (top panel), $\beta$ (middle panel), and UV excess (bottom panel) as a function of \nh\ value for 
1ES\,0229+200, PKS\,0548-322, 1ES\,1741+196, 1ES\,2344+514. }
\label{figure:parameters}
\end{figure*}

Figure\,\ref{figure:parameters}   presents comparison of $\alpha$,   $\beta$, and UV excess as a function of  \nh\  for 1ES\,0229+200, PKS\,0548-322, 1ES\,1741+196, and 1ES\,2344+514.
The figures illustrate the strong dependence of $\alpha$,   $\beta$, and UV excess parameters for all blazars. 
The $\alpha$ and $\beta$ parameters of the model change significantly within changes of the \nh\ value, and this also generates different UV excesses.  

Accoring to the limitations of the values of $\beta$ and UV excess, we can conclude that the values of \nh\ must be limited to:
 13.3$\cdot$10$^{20}$\,cm$^{-2}$, 3.1$\cdot$10$^{20}$\,cm$^{-2}$, 10.5$\cdot$10$^{20}$\,cm$^{-2}$, and 17.4$\cdot$10$^{20}$\,cm$^{-2}$ for 1ES\,0229+200, PKS\,0548-322, 1ES\,1741+196, and 1ES\,2344+514, respectively.

\noindent

\section{Summary} 
\label{summary}
The X-ray spectra of blazars in the energy range of 0.3-10\,keV are commonly well explained using a  power-law or curved power-law model. 
Both these forms can well describe the particle distribution responsible for the emission observed in a given energy range \citep[e.g.][]{Tramacere2007,  Massaro2008}. 
The curvature parameter reported for X-ray spectra of blazar, ranging between 0.1-0.9,  illustrates a variety of spectral properties of blazars'  X-ray spectra \citep [see, e.g.][]{Massaro2004, Tramacere2007, Wierzcholska2016}.
The curved \po\ model is successfully used for all subclasses of blazars, including FSRQs, LBLs, IBLs, and HBLs \citep[e.g.,][]{Wierzcholska2016}.

However, a concave curvature seen in the soft X-ray range can be either an intrinsic feature of the blazar or can be caused by the absorption that is not included in the spectral model.
Usually, using only X-ray observations of the blazars, it is not possible to judge either intrinsic curvature or a  need for an additional absorbing component is a more plausible scenario that can explain the spectral shape. 
One way to verify the origin of spectral curvature is dedicated observations that confirm the existence of gas that could be intrinsic to the blazar or located somewhere between the source and an observer.
Absorption features in the X-ray regime, around 0.5-0.6keV  have been detected in the case of few blazars including: PKS\,2155-304 \citep{Canizares_2155, Madejski_2155}, H\,1426-428 \citep{Sambruna_1426}, 3C\,273 \citep{Grandi_273}, PKS 1034-293 \citep{Sambruna}, PKS\,0548-322 \cite{Sambruna_1998}.
Also, absorption features have been reported in regime of 0.15-0.20\,keV for PKS\,2155-304 \citep{Koenigl_2155} and Mrk\,421\cite{Kartje_421}.

Alternatively, simultaneous multiwavelength data covering a significant part of the broadband spectral energy distribution can be used to get more constraints on the X-ray spectrum and absorption in the soft X-ray range.

We use the models characterized with host galaxy, curvature of the X-ray spectrum, absorption and UV excess in order to describe broadband emission observed from infrared frequencies up to the X-ray band of five extreme HBLs: 1ES\,0229+200, PKS\,0548-322, RX\,J1136+6737, 1ES\,1741+196, 1ES\,2344+514.

We investigate whether curvature that is seen in the X-ray spectra in the energy range of 0.3-10\,keV is an intrinsic feature or caused by the absorption.

Our findings are summarized as follows:

\begin{itemize}

 \item  In the case of four blazars: 1ES\,0229+200, PKS\,0548-322,  1ES\,1741+196, 1ES\,2344+514 prefered spectral characteristic in the X-ray range is a \lp\ model, while for RX\,J1136+6737 a \po\ model is favourable. For all four blazars described with a \lp\ model, spectral fits with the LAB \nh\ value are characterized with a UV excess. 
 If such excess is a real feature of the broadband spectral energy distribution, it can be interpreted either as an additional component such as a blue bump or by unaccounted thermal emission from the AGN.  The latter case, however, seen to be less plausible since it then should be at the order of one or even two magnitudes above the host galaxy template.

  \item In several works focusing on X-ray spectra authors have discussed also problem of proper \nh\ correction \citep[e.g.][]{Acciari_wcom, Wierzcholska2016}.
By using multiwavelength data, we could constrain the X-ray spectrum with all its aspects, including proper correction for hydrogen column absorption. 
 We then conclude that in the case of four sources mentioned, additional absorption,  than the values quoted in by \cite{Kalberla2005}, is needed to explain the spectral properties of these targets in the energy range of 0.3-10\,keV. In the case of 1ES\,0229+200, PKS\,0548-322, 1ES\,1741+196 \nh\ values proposed in this work are higher than previously reported. 
 This column density absorption can be either intrinsic to the source or caused by Galactic absorption in addition to the atomic neutral hydrogen from \cite{Kalberla2005}.
 In the case of PKS\,0548-322, \cite{Sambruna_1998} have reported the possibility of a presence of circumnuclear ionized gas that could explain the need for an additional absorption.
The upper limit of the intergalactic gas luminosity is about 15$\%$ of the BL Lac luminosity.

 \item The \lp\ model is commonly used to describe the X-ray spectra of blazars in the energy regime of 0.3-10\,keV. 
 By using extrapolated X-ray data together with host galaxy template and multiwavelength observations of blazars,  these preferred spectral shapes were confirmed. 
Albeit, curvature seen in the spectra of 1ES\,1741+196, 1ES\,2344+514 is negligible within the uncertainties.
 We then conclude that only in the case of the blazar PKS\,0548-322 the intrinsic spectral curvature is confirmed. 
 This suggests that for PKS\,0548-322 particle population responsible for the synchrotron emission observed should be assumed to be curved as well.

 \item  For four blazars with the UV excess in the broadband SED while using the LAB value of \nh. The excess is no longer present in the spectrum, in the case when a higher amount of the absorption is used. 

 The component needed to explain UV excess  is consistent with the Galactic column density value from the survey by \cite{Willingale13} (5$\%$-10$\%$ difference only) and up to 60$\%$ higher than the amount quoted by \cite{Kalberla2005}.
 The uncertainties of the LAB survey are estimated as 2$\%$.

 \item  In the case of \one\ existence of the cut-off in the UV regime is not confirmed. 
 Such a feature has been previously reported by \cite{Kaufmann2011}. 
 However, the authors constrained the synchrotron peak in the SED of \one\ with poorer multiwavelength coverage, as presented in this work. 
\end{itemize}

\section*{Acknowledgements}
A.W. is supported by Polish National Agency for Academic Exchange (NAWA). 
The plots presented in this paper are rendered using Matplotlib \citep{matplotlib}.
This research has been supported by BMBF through Verbundforschung
Astroteilchenphysik, grant number 05A17VH5.

\bibliographystyle{mn2e_williams}
\bibliography{hosts}

\end{document}